\documentclass{cup-hpl-custom}
\usepackage{hyphenat} 
\usepackage{lipsum}
\usepackage[numbers, sort&compress]{natbib} 
\usepackage{doi} 
\usepackage{comment}
\usepackage{subcaption}

\renewcommand{\deg}{\ensuremath{\mathring{\;}}} 

\begin{document}

\newtheorem{theorem}{Theorem}

\shorttitle{High-rep all-reflective optical guiding and LWFA in helium with an off-axis axicon}
\shortauthor{J. Šišma et al.}

\title{High-repetition-rate, all-reflective optical guiding and electron acceleration in helium using an off-axis axicon}

\useAuthorBlocks  

\authorWithAffils{}{%
Jiří Šišma\textsuperscript{1,2}\corresp{\email{jiri.sisma@eli-laser.eu}},
Michal Nevrkla\textsuperscript{1,2},
Filip Vitha\textsuperscript{1,2},
Sebastian Lorenz\textsuperscript{1},
Illia Zymak\textsuperscript{1},
Alžběta Špádová\textsuperscript{1,2},
Andrea Kollárová\textsuperscript{1,2},
Matěj Jech\textsuperscript{1,3},
Alexandr Jančárek\textsuperscript{1,2},
Davorin Peceli\textsuperscript{1},
Carlo M. Lazzarini\textsuperscript{1},
Leonardo V. N. Goncalves\textsuperscript{1},
Gabriele M. Grittani\textsuperscript{1},
Sergei V. Bulanov\textsuperscript{1}
}{%
\textsuperscript{1}\textit{ELI Beamlines Facility, The Extreme Light Infrastructure ERIC, Za Radnicí 835, Dolní Břežany 25241, Czech Republic}\\
\textsuperscript{2}\textit{Faculty of Nuclear Sciences and Physical Engineering, Czech Technical University in Prague, Břehová 7, 11519 Prague, Czech Republic}\\
\textsuperscript{3}\textit{Faculty of Information Technology, Czech Technical University in Prague, Thákurova 9, 16000 Prague 6, Czech Republic}%
}

\authorWithAffils{}{%
Jaron E. Shrock\textsuperscript{4},
Ela Rockafellow\textsuperscript{4},
Ari J. Sloss\textsuperscript{4},
Bo Miao\textsuperscript{4},
Scott W. Hancock\textsuperscript{4},
Howard M. Milchberg\textsuperscript{4,5}
}{%
\textsuperscript{4}\textit{Institute for Research in Electronics and Applied Physics and Department of Physics, University of Maryland, College Park, Maryland 20742, USA}\\
\textsuperscript{5}\textit{Department of Electrical and Computer Engineering, University of Maryland, College Park, Maryland 20742, USA}%
}

\begin{abstract}
We present recent results on high-power guiding and laser wakefield acceleration (LWFA) in the ELBA beamline at ELI Beamlines, using the L3-HAPLS laser system (13~J, 30~fs, 0.2~Hz). By employing self-waveguiding in a 20~cm plasma channel in helium, we achieved stable acceleration of electron beams to energies of $\sim 5$~GeV. A novel all-reflective optical setup, including an off-axis reflective axicon, enabled efficient acceleration at 0.2~Hz and guiding at repetition rates up to 3.3~Hz. This compact single laser, single compressor implementation of plasma channels for electron acceleration stabilizes electron pointing and enhances energy gain without requiring modifications to the laser system, paving the way for broader adoption of the technology across user facilities.
\end{abstract}


\maketitle

\section{Introduction}
Laser wakefield acceleration (LWFA) enables compact generation of high-energy electron beams \cite{Tajima1979}, which are driving advances in several fields. LWFA electron beams are being used to develop compact free-electron lasers (FELs) that produce ultrashort, high-brilliance X-ray and infrared radiation \cite{Galletti2024, Wang2021, Andre2018}. These electron beams also serve as sources for secondary particles and radiation, including high-quality positron beams \cite{Sarri2013, Sarri2015}, high energy photons \cite{Corde2013, Yan2017, Mirzaie2024, Cole2018}, and even muons \cite{Dreesen2014, Zhang2025, Terzani2025}, enabling applications in fundamental physics \cite{Poder2018, Gonoskov2022}, such as probing techniques \cite{Zhang2017, Wan2023, Russell2023}, as well as in radiotherapy \cite{Labate2020, Guo2025, Zhou2025}. Furthermore, the extremely high acceleration gradients achievable with LWFA---$10^3$ times higher than in conventional RF accelerators \cite{Degiovanni2016} allowing acceleration to multi-GeV energies in sub-meter scale plasma \cite{Leemans2006, Clayton2010, Miao2022, Aniculaesei2024}---open the door to next-generation high-energy particle colliders, potentially reducing the size and cost of future TeV-scale accelerators \cite{Schroeder2010}. 

The electron energy gain in LWFA is primarily limited by laser diffraction, pulse depletion and dephasing. To overcome diffraction and maximize energy gain, several guiding  techniques have been developed. One approach is self-guiding, where an ultra-high intensity laser pulse creates its own plasma channel via relativistic and ponderomotive self-focusing, balancing diffraction and enabling propagation over several Rayleigh lengths \cite{Clayton2010, Kim2013, Aniculaesei2024}. However, it is a nonlinear process highly dependent on plasma density, which can limit controllability, stability and efficiency. In contrast, pre-formed plasma channels, created in advance using methods such as optical-field ionization (OFI) \cite{Morozov2018, Lemos2018, Shalloo2018, Smartsev2019, Shalloo2019, Miao2020, Feder2020, Picksley2020a, Picksley2020b, Miao2022, Shrock2022, Shrock2024, Miao2024, Picksley2024,  Rockafellow2025, Rockafellow2025b} or capillary discharge \cite{Bobrova2001, Butler2002, Leemans2006, Karsch2007, RowlandsRees2008, Gonsalves2019}, provide a stable, low-density guiding structure without reliance on nonlinear propagation effects. In LWFA, this enables propagation over much longer distances (tens of centimeters to meters) and supports higher electron energies \cite{Gonsalves2019, Miao2022, Picksley2024, Rockafellow2025} by enabling acceleration at lower plasma densities. Pre-formed channels also offer better control over the plasma density profile, matched laser spot size, and overall stability, reducing shot-to-shot fluctuations and improving beam quality \cite{Shrock2024, Tripathi2025}.

The first plasma waveguides were demonstrated in high-density plasma using inverse Bremsstrahlung heating \cite{Durfee1993, Durfee1995}. This heating mechanism is inefficient and unsuitable for the low plasma densities required for multi-GeV LWFA \cite{Miao2022, Picksley2024, Rockafellow2025}, as it cannot achieve on-axis densities below $10^{17}$ cm$^{-3}$. However, extending the method to use density-independent optical-field ionization as the heating mechanism \cite{Corkum1989, Lemos2018, Shalloo2018, Smartsev2019, Miao2020, Feder2020}, together with auxiliary ionization of the neutral shock to form the cladding \cite{Morozov2018, Miao2020, Feder2020, Picksley2020b, Shrock2022}, has enabled the production of highly confining, meter-scale plasma waveguides with on-axis densities below $10^{17}$ cm$^{-3}$. These free-standing waveguides are immune to laser damage, longitudinally tunable \cite{Shrock2024, Tripathi2025}, and can reliably reach the low densities required for high-energy acceleration \cite{Miao2022, Picksley2024, Rockafellow2025} using gas jet technology \cite{Lorenz2019, Shrock2022, Miao2025}. In contrast, capillary-based waveguides are limited by their fixed geometry and susceptibility to laser-induced damage \cite{Pieronek2020}, and inability to reach densities as low as $\sim 1 \times 10^{17}~\mathrm{cm^{-3}}$\cite{Gonsalves2019}. Electron energies of up to 8 GeV have been achieved using capillary discharge waveguides enhanced with auxiliary laser heating \cite{Gonsalves2019}.

Advances in laser technology have enabled the use of hydrodynamically expanded OFI plasma columns to form plasma waveguides, allowing for longer acceleration lengths and improved control over plasma profiles \cite{Morozov2018, Shalloo2018, Picksley2020b, Miao2020, Feder2020}. However, at low target gas density, a single femtosecond channel-forming pulse may not create the core and adequately confining cladding of a plasma waveguide structure by itself. Instead, it ionizes a plasma column causing a hydrodynamic expansion with a plasma ``core'' on axis and neutral density shock expanding outwards. A secondary pulse is required to ionize the neutral density shock to form the ``cladding'' and establish the refractive index required for guiding \cite{Morozov2018, Miao2020, Feder2020, Picksley2020b}. Two main techniques have been developed: the ``2-Bessel'' method \cite{Miao2020}, which uses a secondary higher-order Bessel beam to ionize the neutral shock of the OFI channel, allowing guiding of a high or low intensity pulse, or ``self-waveguiding'' technique \cite{Morozov2018, Feder2020, Picksley2020b}, which simplifies the process by using the leading edge of an intense pulse to ionize the neutral shock, forming the waveguide structure for itself as it propagates, simultaneously driving the wake. Both methods have recently been implemented with phase front correction \cite{Miao2022oe, Miao2022, Turner2022} to produce high-fidelity Bessel beams, which are crucial for stable and efficient plasma channel formation. Single-stage plasma accelerators have achieved electron energies up to 10 GeV, using nanoparticle-assisted self-guided LWFA with 130~J on target \cite{Aniculaesei2024} and self-waveguiding of the LWFA driver pulse with 18 J \cite{Rockafellow2025} and 21 J \cite{Picksley2024} on target.

\begin{figure*}[t]
\centering
\includegraphics[width=\textwidth]{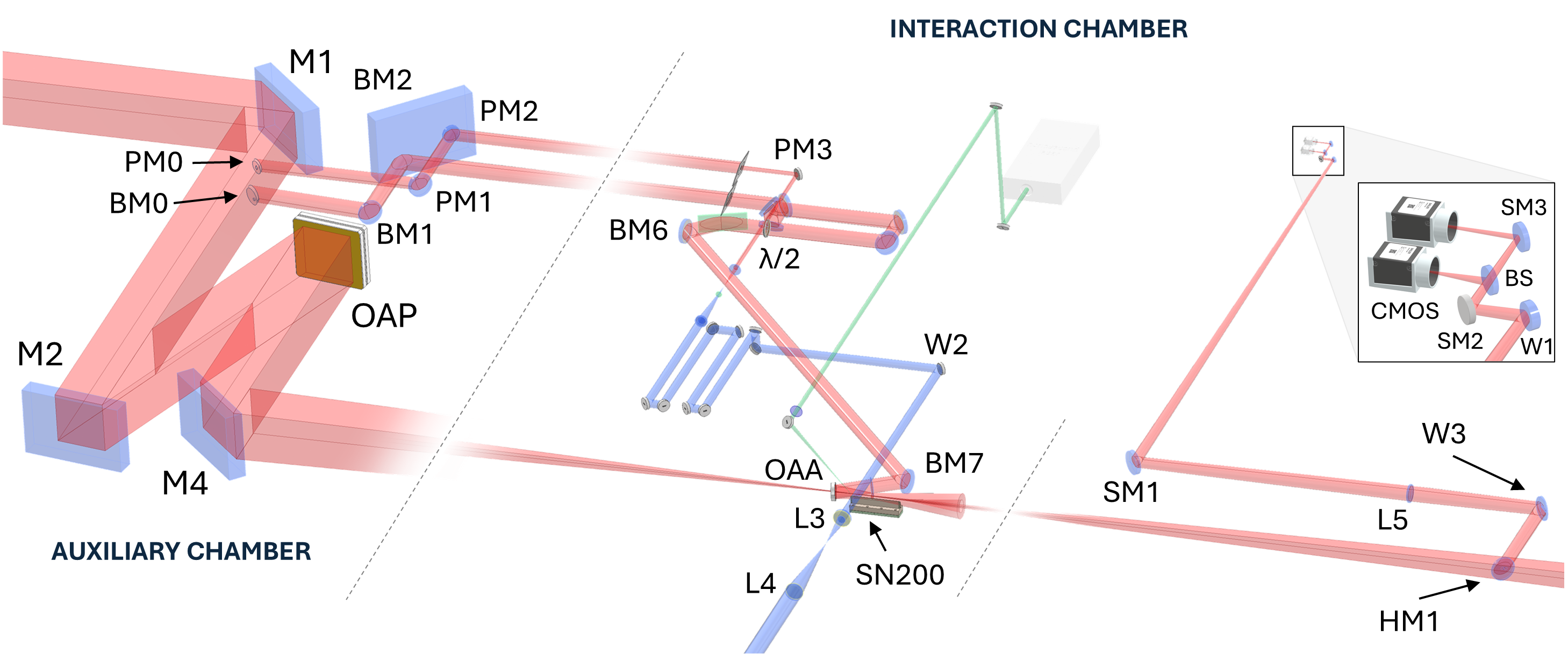}
\caption{\textit{Experimental setup overview.} The laser beam enters the auxiliary chamber from the left and is reflected by mirror M1. Two pick-off mirrors split small portions of the pulse to form the channel-forming and probe beams, while the main pulse continues as the LWFA drive beam. The channel-forming beam is routed through the auxiliary chamber and directed into the interaction chamber toward the off-axis axicon (OAA) positioned above the gas jet SN200, where it generates the plasma channel. The probe beam is guided through its delay line and diagnostic path, and the drive beam is focused by the off-axis parabola (OAP) into the interaction region for laser wakefield acceleration. The high-power focal-spot and guided-beam diagnostics are shown on the right side of the interaction chamber and include a mirror with a central aperture (HM1), an uncoated wedge (W3), an imaging lens (L5), and a mirror (SM1) that directs the attenuated pulse out of the chamber to CMOS cameras. The green beam path indicates a separate nanosecond laser system used in independent experiments to test alternative injection mechanisms. A detailed description of the setup is provided in the \textit{Experimental setup} section below.}
\label{fig:setup}
\end{figure*}

We present a new modification of the self-waveguiding technique that employs post-compressor splitting of the channel-forming pulse from an initially square transverse profile (super-Gaussian in each dimension) femtosecond laser pulse. This approach primarily relies on reflective optics, thereby avoiding heat deposition in transmissive components and preventing beam distortions caused by B-integral effects in short pulses. For the first time in high-power LWFA experiments, we implemented an off-axis reflective axicon \cite{Boucher2018}, offering straightforward alignment and reduced risk of laser damage compared to traditional on-axis back-reflective configurations \cite{Miao2020, Feder2020}. The all-reflective setup preserved the compressed pulse duration after the axicon and enabled the formation of plasma channels in helium, which requires laser intensities above $9 \times 10^{15}$ W$/$cm$^2$ for full ionization \cite{Corkum1989, Tong2005}. Previous studies have shown that the plasma channel properties of OFI-generated plasmas can be maintained even at kHz repetition rates \cite{Alejo2022}, confirming the suitability of this approach for future high-repetition-rate accelerators and colliders. In our experiments, we implemented the self-waveguiding LWFA scheme of Ref.\citenum{Miao2022}, where 5 GeV electron acceleration was previously demonstrated in hydrogen. Using an all-reflective geometry, we achieved stable optical guiding of high-intensity laser pulses in a 3 cm plasma channel at 3.3 Hz, and, for the first time in helium, we demonstrated electron acceleration to energies up to 5 GeV in a 20 cm plasma channel. The use of helium as the working gas provides a safer and more easily pumped alternative to hydrogen, which was commonly used in similar experiments \cite{Miao2022, Shrock2024, Picksley2024, Rockafellow2025} due to its lower ionization threshold. This work demonstrates that the proposed approach is both adaptable and efficient for laser user facilities where laser system modification is limited by operational constraints, while enabling maximized electron energy gain compared to conventional self-guiding methods. Owing to its single-compressor implementation, the scheme is well suited for future compact muon sources.

\begin{figure*}[t]
\centering
\includegraphics[width=\textwidth]{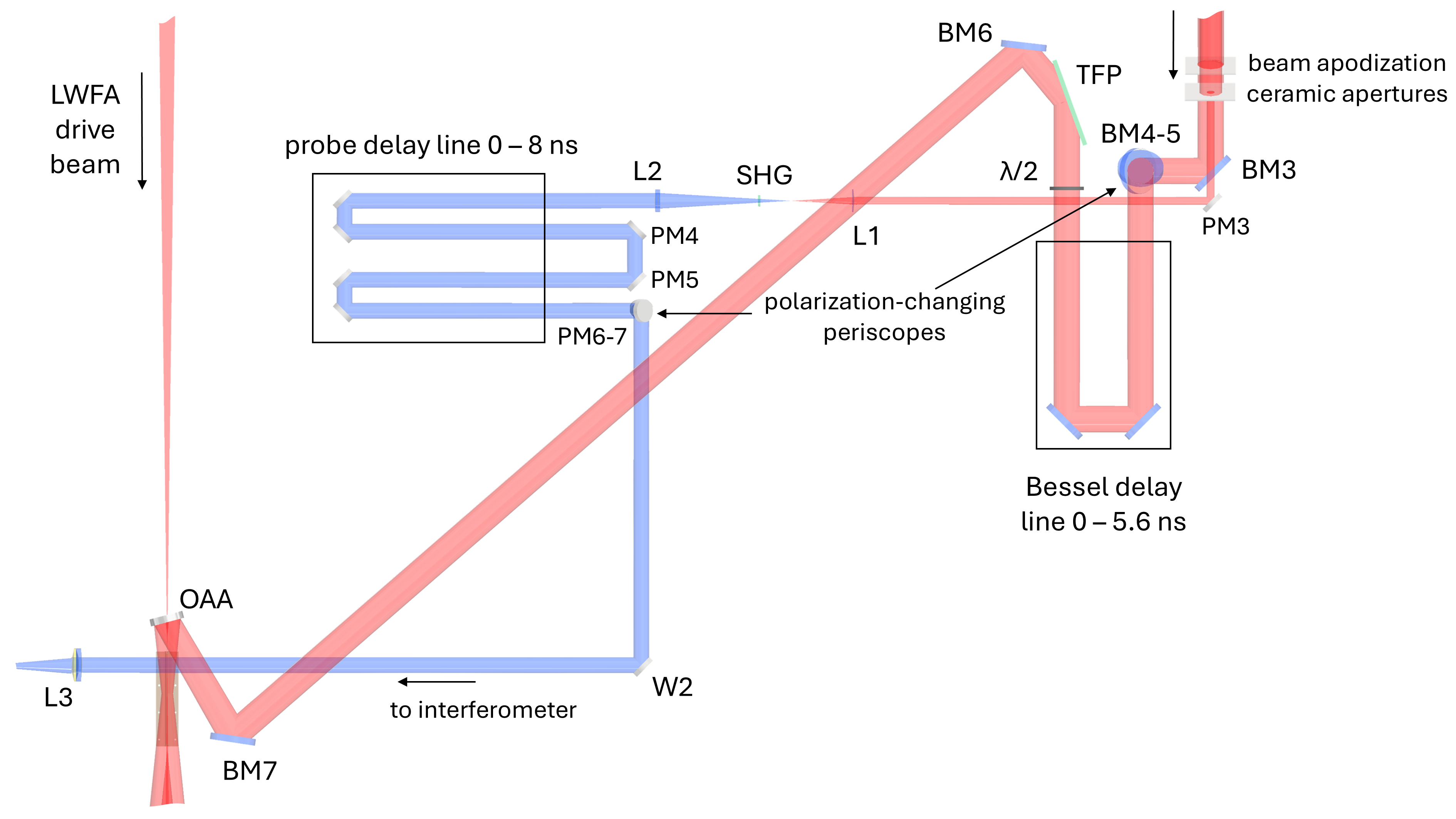}
\caption{\textit{Schematic layout of the interaction chamber setup for the self-waveguided LWFA experiment.} The final section of the optical system is shown, starting from mirror BM3, where the channel-forming (Bessel) beam enters the chamber, and including all subsequent mirrors BM3–BM7, the periscope, attenuator (half-wave plate, $\lambda /2$, and thin-film polarizer, TFP), and delay-line mirrors leading to the off-axis axicon (OAA) positioned above the 20 cm gas jet SN200. The LWFA drive beam passes through the axicon’s central aperture and interacts with the plasma channel formed by the Bessel beam, while the probe beam is directed from its apodization aperture via mirror PM3 to the telescope (lenses L1 and L2) with an SHG BBO crystal, then through a delay line toward the interferometer for diagnostics. A detailed description of the setup is provided in the \textit{Experimental setup} section.}
\label{fig:setup-interaction}
\end{figure*}

\section{Experimental setup} \label{sec:experimental_setup}

The experiments were performed at the ELBA (ELectron Beam Accelerator) beamline \cite{Lazzarini2019} using the L3-HAPLS Ti:sapphire laser system at the ELI Beamlines facility in the Czech Republic. In these experiments, the laser delivered linearly polarized, 30 fs pulses with peak power of 400 TW at central wavelength 810 nm, providing up to 13 J of energy at 0.2~Hz repetition rate, or up to 8 J at 3.3 Hz. The beam had a rectangular super-Gaussian profile with dimensions of $214 \times 214$ mm.

In Figure~\ref{fig:setup}, the laser beam enters the auxiliary chamber from the left and is first reflected by a high-reflectivity (HR) dielectric mirror\cite{Willemsen2022} (M1). Two pick-off mirrors are then used to split small portions of the pulse to form two auxiliary beams: a channel-forming beam and a probe beam.

The main portion of the beam continues to an HR mirror (M2) and a 20$\deg$ off-axis parabola (OAP) with a 10~m focal length. The reflected beam is directed by mirror M4 through the central aperture of the off-axis axicon, focusing above the gas jet. Mirror M4 also enables fine alignment between the channel entrance and the drive beam without requiring OAP adjustments, thereby preventing focal spot distortion and ensuring stability during extended operation.

The channel-forming (Bessel) beam is separated by a 45$\deg$ silver-coated elliptical mirror (BM0) with a 66 mm minor diameter, extracting up to 1.1 J from the initial 13 J pulse. This beam is subsequently guided by a dielectric 4-inch mirror (BM1) and an HR mirror (BM2) into the interaction chamber, connected to the auxiliary chamber by a 6.2 m long beam-transport tube. Within this section, the beam can be apodized using ceramic apertures to achieve the desired diameter and suppress diffraction. All mirrors BM0-BM2 are motorized to allow smooth, remote alignment over long distances, ensuring stability and reproducibility throughout the experiment.

In the remaining beamline (see Figure~\ref{fig:setup-interaction}), enhanced silver-coated mirrors (BM3–BM7 and delay-line mirrors) were employed. The single compressor scheme and size of the vacuum chambers constrain the delay to a maximum of 5.6~ns between the LWFA drive pulse and the Bessel beam, adjustable via an 840 mm single-pass delay-line stage. The initially linearly p-polarized channel-forming beam passes through a polarization-changing periscope (BM4–BM5) that rotates its polarization by 90$\deg$ to work in a reflective attenuator configuration designed for s-polarization. A beam attenuator consisting of a half-wave plate ($\lambda/2$) and a reflective femtosecond thin-film polarizer (TFP) was implemented to enable precise control of the Bessel-beam energy \cite{Miao2024, Rockafellow2025}, providing flexibility in tuning the plasma waveguide parameters. The attenuator could be bypassed, allowing operation with only reflective components in the beam path suitable for high-repetition rate operation. Data presented in this paper were taken without the attenuator installed.

Finally, mirrors BM6 and BM7 were used to align a 3-inch reflective 30$\deg$ off-axis axicon (OAA) with a 3$\deg$ base angle ($\approx 105$ mrad approach angle) and a 7 mm central aperture, which allows the LWFA drive beam to couple into the pre-formed density structure above a 20~cm-long supersonic slit-nozzle single-valve gas jet SN200. It is a supersonic slit nozzle system developed at ELI Beamlines \cite{Lorenz2019} for long laser--plasma-driven electron acceleration schemes. The nozzle is designed to produce a 20~cm-long region of uniform gas density above the slit, with minimal density variations along the flat-top profile. The system incorporates a gas reservoir and a set of flow deflectors positioned upstream of the nozzle throat to maintain a stable and uniform density distribution, even when operated with a single gas valve connected to the nozzle assembly. This nozzle system differs from previously reported designs \cite{Shrock2022, Miao2025, Rockafellow2025} in that the gas is supplied through a single valve. This configuration significantly simplifies implementation of the system and allows the target to be operated using readily available commercial driver hardware. The trade-off, however, is a comparatively long valve-opening time ($\approx$ 10~ms), which is required for the gas density profile to fully develop prior to the arrival of the laser pulse.

Similarly, the probe beam was extracted from the main laser beam using a 2-inch elliptical, protected silver-coated mirror (PM0). It was then directed by a 4-inch enhanced silver-coated mirror (PM1) and a 3-inch dielectric mirror (PM2) toward the interaction chamber, where it was apodized using a ceramic aperture. The beam was subsequently reflected by a 2-inch protected silver-coated mirror (PM3) into a telescope consisting of a thin lens (L1) and an achromatic lens (L2), transmitting through a beta barium borate (BBO) crystal for second-harmonic generation (SHG) at 800 nm (see Figure~\ref{fig:setup-interaction}). For the experimental results discussed in this paper, the fundamental (first harmonic) probe was used exclusively as the diagnostic beam.

The probe beam then passed through a double-pass, 600~mm-long delay line stage, enabling precise timing of the probe pulse with respect to the drive pulse from –2~ns to +6~ns. This allowed temporal scanning of the plasma channel evolution from 0 to 8 ns by adjusting the Bessel beam delay.

Finally, the probe beam polarization was rotated using a polarization-changing periscope (PM6–PM7), enabling the filtering of p-polarized scattered driver light, and then transmitted through an uncoated wedge (W2) into the interaction region. The transmitted beam was focused by a 3-inch, 200~mm-focal-length achromatic convex lens (L3), forming the first element of a Keplerian telescope that guided the beam out of the chamber through a large-aperture optical window. The beam was then collimated by another 3-inch, 400~mm-focal-length achromatic convex lens (L4) before entering a folded-wave Mach–Zehnder interferometer, where the image was focused onto a complementary metal oxide semiconductor  (CMOS) camera with a $2\times$ magnification.

The scanning interferometer system enables spatial scanning of the probe beam along the gas jet through synchronized motorized control of the wedge W2, the first lens L3, and the entire out-of-chamber interferometer assembly, including lens L4, allowing independent and repeatable positioning along the interaction region.

For the post-interaction diagnostics, we employed a high-power focal-spot and guided-mode diagnostic system consisting of a 4-inch drilled uncoated ultraviolet fused silica (UVFS) mirror (HM1) with a grinded back surface, a 4-inch uncoated wedged (W3) mirror, a 3-inch imaging lens (L5) with a 2000~mm focal length, and a 4-inch protected silver-coated mirror (SM1). Outside the chamber, the beam was reflected by a 1-inch uncoated wedge (W1) and directed by a silver mirror (SM2) toward the first beamsplitter (BS), which sent one portion to a CMOS camera imaging the focal plane. The second image was focused onto another CMOS camera monitoring the guided mode at the exit plane of the waveguide. The cameras used were spectrally sensitive in the visible and near-infrared (near-IR) range up to 1100~nm. Each UVFS reflection at $45 \deg$ attenuates the p-polarized beam to $0.66~\mathrm{\%}$ of the initial incident power at 800 nm. With three uncoated UVFS reflections, the beam is attenuated by a factor of $3 \times 10^{-7}$. To prevent detector saturation, absorptive neutral-density filters were placed in front of the cameras for additional attenuation with the total optical density ranging from 3 to 6, corresponding to attenuation factors from $10^{-3}$ to $10^{-6}$.

The accelerated electron beam passed through a 10 mm central aperture in the first mirror HM1, positioned 3 m downstream from the nozzle exit (corresponding to a 3~mrad divergence). After exiting the chamber through a 1 mm-thick aluminum flange, it propagated toward a 1 mm tungsten slit, composed of 6 cm-thick blocks, placed directly in front of a 50 cm-long slit electromagnet operating at a uniform magnetic field of 0.6~T. A LANEX screen was mounted at the exit of the electromagnet and imaged by a CMOS camera equipped with a camera lens and a 546~nm bandpass filter with a 10~nm bandwidth for energy-spectrum diagnostics. The electron spectrometer was calibrated in charge using a conventional linac at energies $6-20$~MeV \cite{Zymak2024}, and correction factors were applied to account for differences in the geometric arrangement of the setup.

We also employed \textit{top-view} and \textit{side-view} fluorescence diagnostics \cite{Shrock2022} consisting of CMOS cameras equipped with camera lenses, motorized neutral-density filter wheels, and a helium line bandpass filter centered at 589~nm with a 10~nm bandwidth. The filter isolated the strong, well-separated helium emission line at 587.6~nm, corresponding to the $3d \rightarrow 2p$ transition following recombination. To prevent laser light from reaching the detectors and to improve signal quality, infrared (IR) cut filters were placed in front of the cameras (doubled IR cut at 645~nm with $<1\mathrm{\%}$ averaged transmission in range of $690-1070$~nm for \textit{side-view} camera and doubled IR cut at 710~nm with $<10\mathrm{\%}$ averaged transmission in range of $740-1200$~nm for \textit{top-view} camera). These diagnostics were used for fine alignment of the plasma column above the gas jet and to observe impurities in both the gas sheet and the Bessel beam intensity distribution. The resulting fluorescence signal represents the combined effects of the Bessel beam intensity profile along the propagation and gas density fluctuations or inhomogeneities.

\begin{figure}[t]
\centering
\includegraphics[width=\columnwidth]{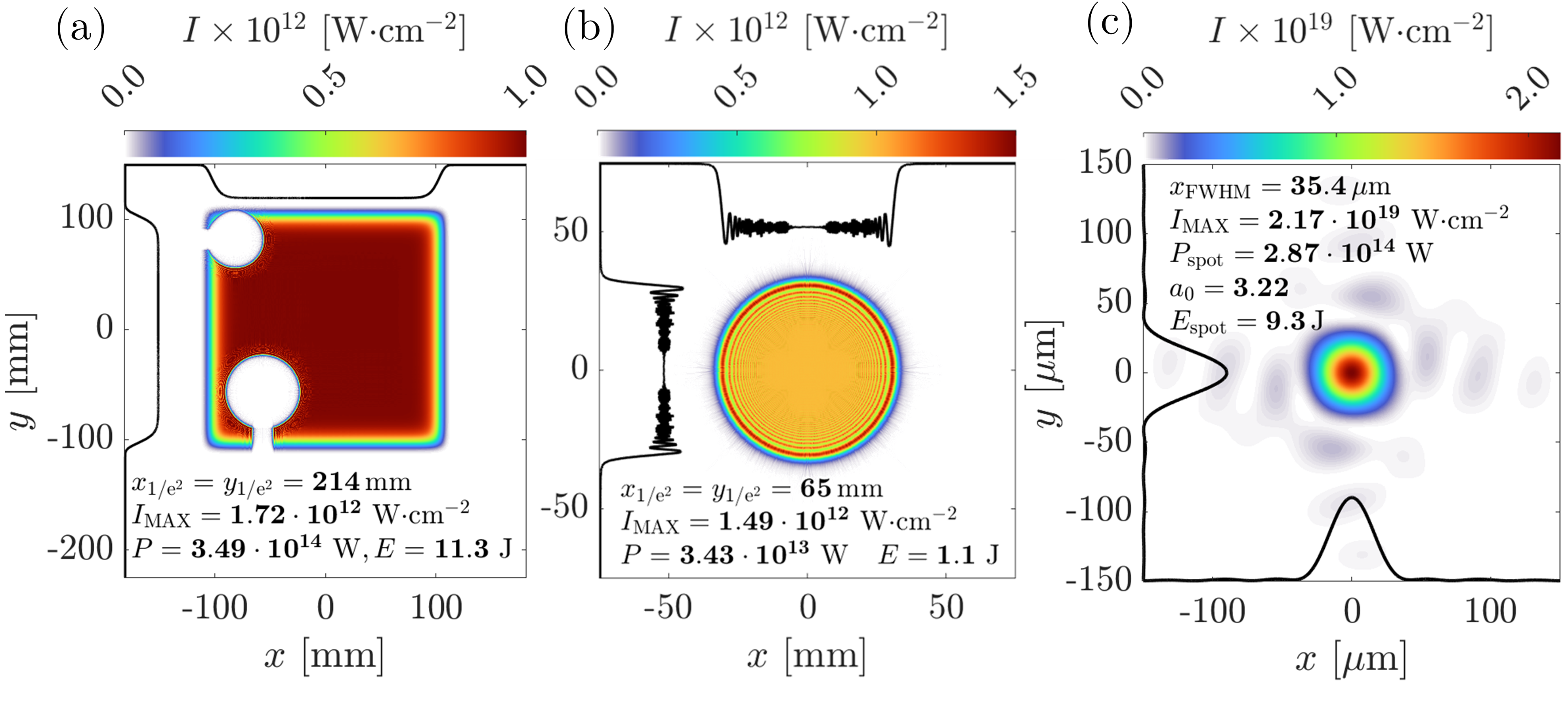}
\caption{\textit{Simulations of laser beam wavefront splitting and focal spot.}
(a) Wavefront of the LWFA driver beam after the pick-offs at the off-axis parabola (OAP). (b) Wavefront of the channel-forming beam after 12 m of free-space propagation at the axicon surface. (c) Focal spot of the LWFA driver beam (a) on target.}
\label{fig:VLsimulations}
\end{figure}

\begin{figure}[t]
\centering
\includegraphics[width=\columnwidth]{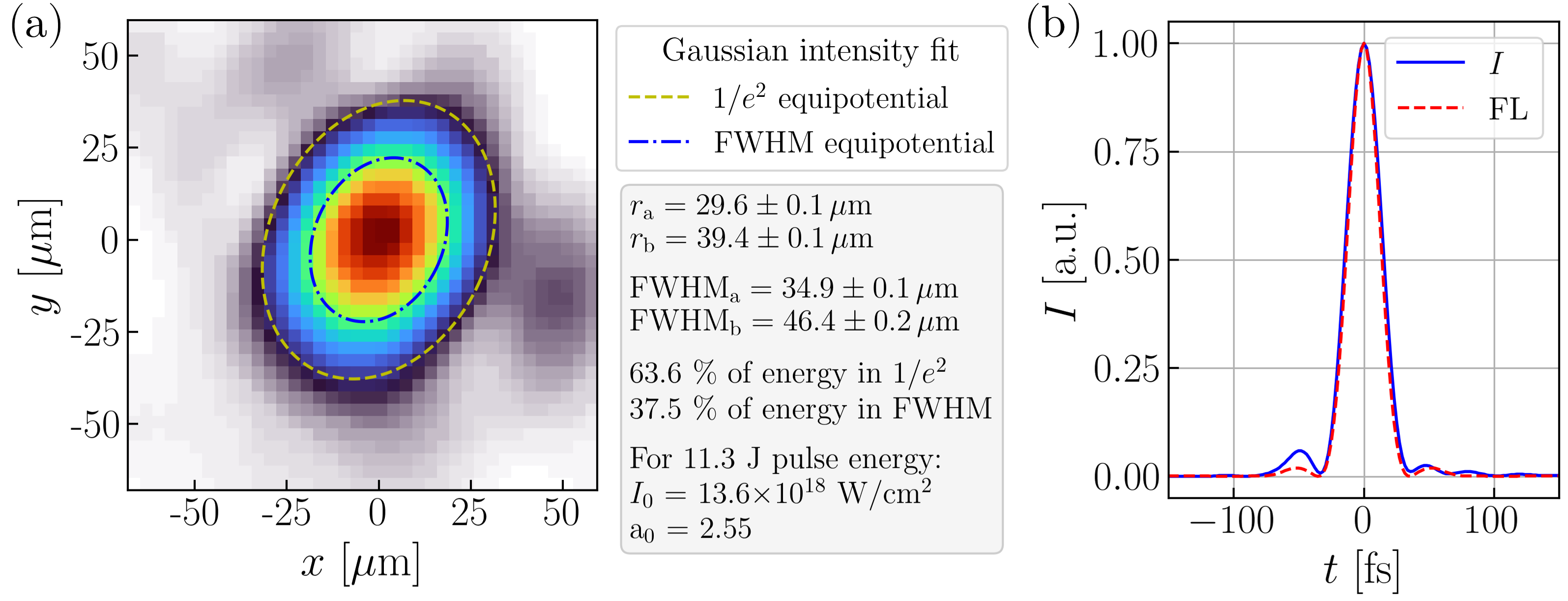}
\caption{\textit{Laser beam measurements.} 
(a)~Low-power focal-spot image and corresponding analysis. 
(b)~Laser pulse characterization using SPIDER measurement, $I$ denotes the measured intensity profile and $FL$ denotes the Fourier-limited laser pulse, i.e. the minimum pulse duration calculated from the measured spectral bandwidth, assuming a flat spectral phase.}
\label{fig:FSmeas}
\end{figure}

\begin{figure*}[h!]
\centering
\includegraphics[width=\textwidth]{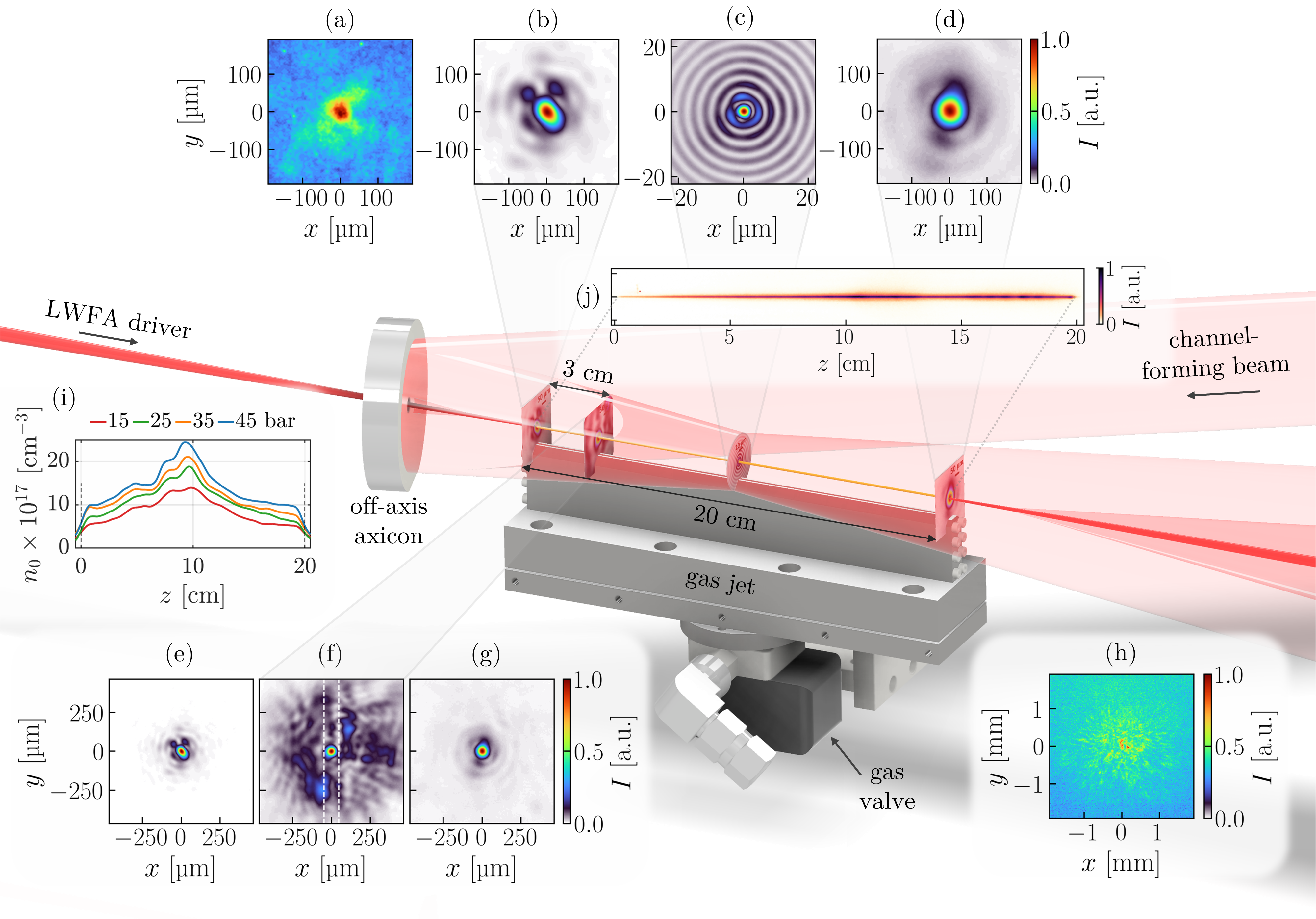}
\caption{\textit{Guiding overview illustrating the interaction between the LWFA drive beam and the channel-forming beam, with the plasma column (yellow) highlighted between the focal plane and the waveguide exit plane above a 20 cm gas jet.}
(a)~High-power focal-spot diagnostic image taken through the gas sheet during active guiding, used for online alignment monitoring.
(b)~High-power focal spot in vacuum, recorded using the focal-spot diagnostic camera.
(c)~Bessel beam focal spot, measured with a CMOS camera with $5\times$ magnification objective.
(d)~Image of a guided mode exiting 20 cm plasma waveguide acquired by the guided-mode diagnostic camera when the driver beam is successfully coupled into the waveguide.
(e)~Zoomed-out high-power focal spot image, same as in (b) for comparison with (f)~Guided-mode exiting 3 cm plasma waveguide using short 3 cm gas jet (see Figure~\ref{fig:hrrguide}), showing leaky modes \cite{Shrock2024} surrounding the guided beam.
(g)~Zoomed-out guided mode exiting 20 cm plasma waveguide, same as in (d).
(h)~Guided-mode image recorded when the driver beam misses the waveguide entrance due to laser-pointing jitter.
(i)~Axial neutral-gas density profiles as a function of backing pressure, measured via helium fluorescence induced by J$_0$ Bessel-beam ionization of the gas sheet 8~mm above the nozzle orifice with a 10~ms valve opening time. The neutral density was calibrated using known helium backfill in the chamber \cite{Shrock2022}.
(j)~\textit{Top-view} fluorescence image of the helium plasma channel generated by the off-axis-axicon--produced J$_0$ Bessel beam. Images (a)-(h) and (j) are individually normalized.}
\label{fig:guiding}
\end{figure*}

\section{Laser splitting and focal spot simulations}

We performed laser beam-splitting and focal-spot simulations for an ideal super-Gaussian beam using the L3-HAPLS laser parameters to compare with the measured focal spots and to estimate the energy distribution in each pulse. Additionally, we simulated the free-space propagation of the Bessel beam to study diffraction-induced distortions after approximately 12~m of propagation. To suppress the resulting diffraction pattern, a serrated aperture can be employed~\cite{Auerbach1994}. All simulations were carried out using \textit{VirtualLab Fusion 7.0.1} \cite{LightTrans_VirtualLabFusion} (fast physical-optics simulation software).

In Figure~\ref{fig:VLsimulations}, we present simulations of the free-space propagation and focal spot of the LWFA drive beam. For a 13~J, 30.4~fs, 10th-order super-Gaussian pulse, the LWFA driver beam carries 11.3~J of energy (Figure~\ref{fig:VLsimulations}(a)), while the channel-forming (Bessel) beam contains 1.1~J. The simulated propagation shows modulation in the near-field of the channel-forming beam intensity due to diffraction effects (Figure~\ref{fig:VLsimulations}(b)). The remaining 0.5~J of energy was allocated to the probe beam. The resulting focal spot corresponding to the wavefront (in Figure~\ref{fig:VLsimulations}(a)) is shown in Figure~\ref{fig:VLsimulations}(c). The expected full width at half maximum (FWHM) is $\approx 35~\mathrm{\mu}\text{m}$, with a normalized vector potential of $a_0 = 3.22$, and a maximum energy in the zeroth diffraction order of 9.3~J, corresponding to $82.3~\mathrm{\%}$ of total energy.

\begin{figure*}[t]
\centering
\includegraphics[width=\textwidth]{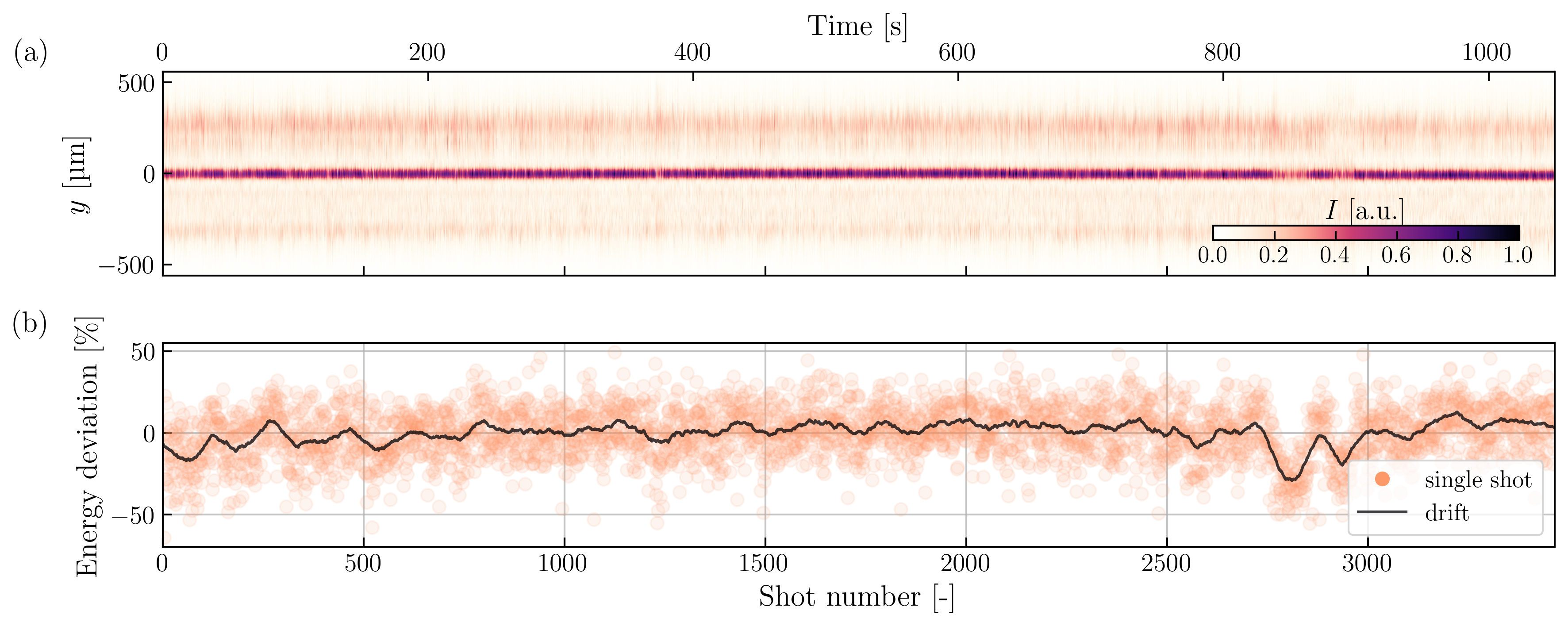}
\caption{\textit{High-repetition-rate guiding results.}
(a) Evolution of the guided mode over approximately 20 minutes of operation at 3.3 Hz. Each shot is represented by a line-averaged intensity profile taken from a fixed region centered on the guided mode.
(b) Evolution of the guided energy deviation within a beam radius (1/e$^2$), together with the energy drift calculated as a moving average over a 60-shot window.
}
\label{fig:hrrguide-data}
\end{figure*}

\section{Experimental results}

We have verified pulse compression in the experimental chamber by employing SPIDER (Spectral Phase Interferometry for Direct Electric-Field Reconstruction) \cite{Iaconis1998} to measure the femtosecond pulse duration, obtaining 30.4~fs (FWHM), as shown in Figure~\ref{fig:FSmeas}(b). The focal spot was characterized in the low-power regime (L3-HAPLS front-end) by mapping the total energy across the full image, resulting in the distribution presented in Figure~\ref{fig:FSmeas}(a). After noise subtraction, the focal spot was fitted with a Gaussian profile, yielding a minor-axis FWHM of $35~\mathrm{\mu}\text{m}$ and a major-axis FWHM of $46~\mathrm{\mu}\text{m}$. The overall energy coupling within the beam radius (1/e$^2$) was estimated to be approximately $64~\mathrm{\%}$ (7.2~J). For an ideal Gaussian intensity profile, the encircled energy within the $\mathrm{1/e^2}$ radius is approximately $86.5~\mathrm{\%}$ of the total energy. The retrieved normalized vector potential was $a_0 = 2.55$. Based on performed simulations, the Strehl ratio (defined as the ratio of measured to simulated $a_0$ in the focal region) was estimated to be 0.79.

\begin{figure}[t]
\centering
\includegraphics[width=\columnwidth]{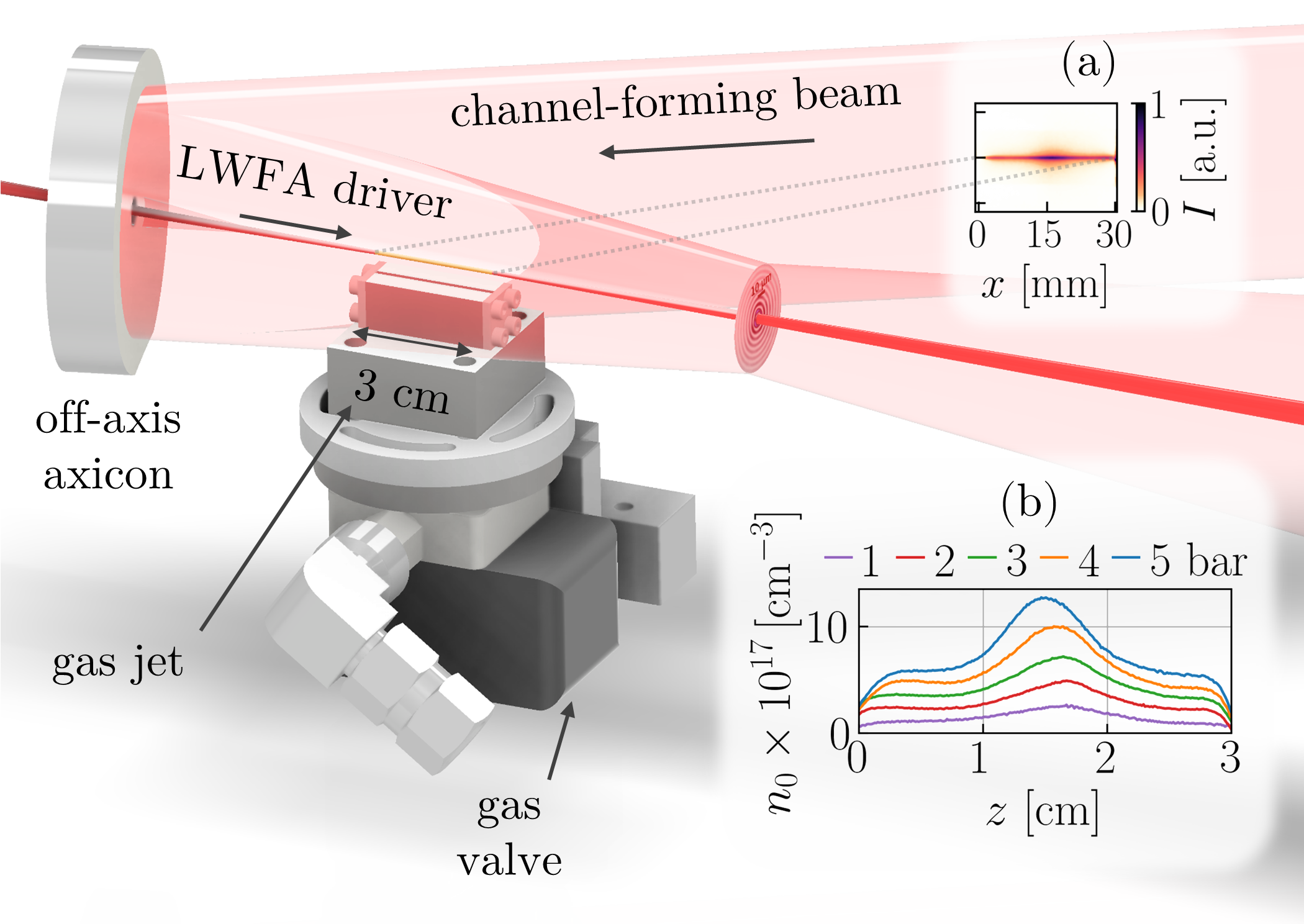}
\caption{\textit{High-repetition-rate guiding overview illustrating the setup used with a 3 cm slit nozzle.}
(a)~Normalized \textit{top-view} fluorescence image of the 3~cm helium plasma channel.
(b)~Axial neutral-gas density profiles measured 5~mm above the nozzle orifice with a 2~ms valve opening time as a function of backing pressure, measured via a \textit{top-view} fluorescence diagnostic.}
\label{fig:hrrguide}
\end{figure}

\subsection{High-repetition-rate, high-power laser guiding}

We characterized high-power laser guiding at 0.2~Hz in a 20~cm helium plasma channel (Figure~\ref{fig:guiding}) and high-repetition-rate guiding at 3.3~Hz using a 3~cm supersonic slit nozzle (Figures~\ref{fig:hrrguide-data} and \ref{fig:hrrguide}). The shorter nozzle and reduced gas opening time were selected to limit the gas load in the interaction chamber, enabling sustained operation at 3.3~Hz while maintaining acceptable vacuum conditions. Figures~\ref{fig:guiding}(a) and \ref{fig:guiding}(b) show high-power focal-spot images recorded through the gas sheet during active guiding and in vacuum at the focal plane, respectively. The transverse intensity profile of the zeroth-order Bessel beam J$_0$ is presented in Figure~\ref{fig:guiding}(c), yielding a first-zero radius of $2.83 \pm 0.01~\mathrm{\mu}\text{m}$ measured with a $5\times$ finite-conjugate objective. The guided mode at the 20~cm exit plane is shown in Figure~\ref{fig:guiding}(d).

Figures~\ref{fig:guiding}(e)-\ref{fig:guiding}(g) provide a direct comparison of the focal-spot and guided-mode distributions on a common spatial scale. The guided-mode image in Figure~\ref{fig:guiding}(f) was obtained with the 3~cm nozzle (corresponding diagnostics shown in Figure~\ref{fig:hrrguide}). In both the 3 and 20~cm configurations, the guided mode becomes more circularly symmetric relative to the incident focal spot (Figure~\ref{fig:guiding}(b)). This behavior is consistent with suppression of higher order mode content via leakage \cite{Antonsen1995, Clark2000, Feder2020} and group velocity walkoff \cite{Shrock2024}. The residual light surrounding the central spot in Figure~\ref{fig:guiding}(f) is attributed to leakage from the waveguide.

High-repetition-rate guiding at 3.3~Hz was demonstrated using the 3~cm nozzle, as shown in Figure~\ref{fig:hrrguide}. Figure~\ref{fig:hrrguide-data}(a) presents the time evolution of the guided mode over $\approx$~20 minutes of operation at 3.3~Hz. Each shot is represented by a line-averaged intensity profile extracted from a fixed region centered on the guided mode (see Figure~\ref{fig:guiding}(f)). The guided energy deviation within a beam radius (1/e$^2$) is shown in Figure~\ref{fig:hrrguide-data}(b), together with the energy drift calculated as a moving average over a 60-shot window. The guided mode remains stable throughout the sequence for a $\approx 250$~TW laser pulse, with no evidence of long-term degradation. A small decrease in guided energy observed after 800 seconds is attributed to a drop in the gas valve backing pressure. The guiding was achieved at a backing pressure of 6~bar, corresponding to an average neutral density of $\approx 9.4\times 10^{17}$~cm$^{-3}$ along the gas jet. The axial neutral-gas density profiles, measured via the \textit{top-view} fluorescence diagnostic, are shown in Figure~\ref{fig:hrrguide}(b). A representative fluorescence image from this diagnostic is shown in Figure~\ref{fig:hrrguide}(a).

\subsection{Electron acceleration in helium}

We achieved electron acceleration in a 20~cm preformed helium plasma channel generated above the supersonic slit nozzle SN200, operated with a single valve and a 10~ms opening time limited by the vacuum pumping system. The neutral-density profiles measured along the gas jet are shown in Figure~\ref{fig:guiding}(i). The observed non-uniformity is attributed to the short opening time, during which the gas does not have sufficient time to develop the flat density distribution expected from the design.
\begin{figure}[h!]
\centering
\includegraphics[width=\columnwidth]{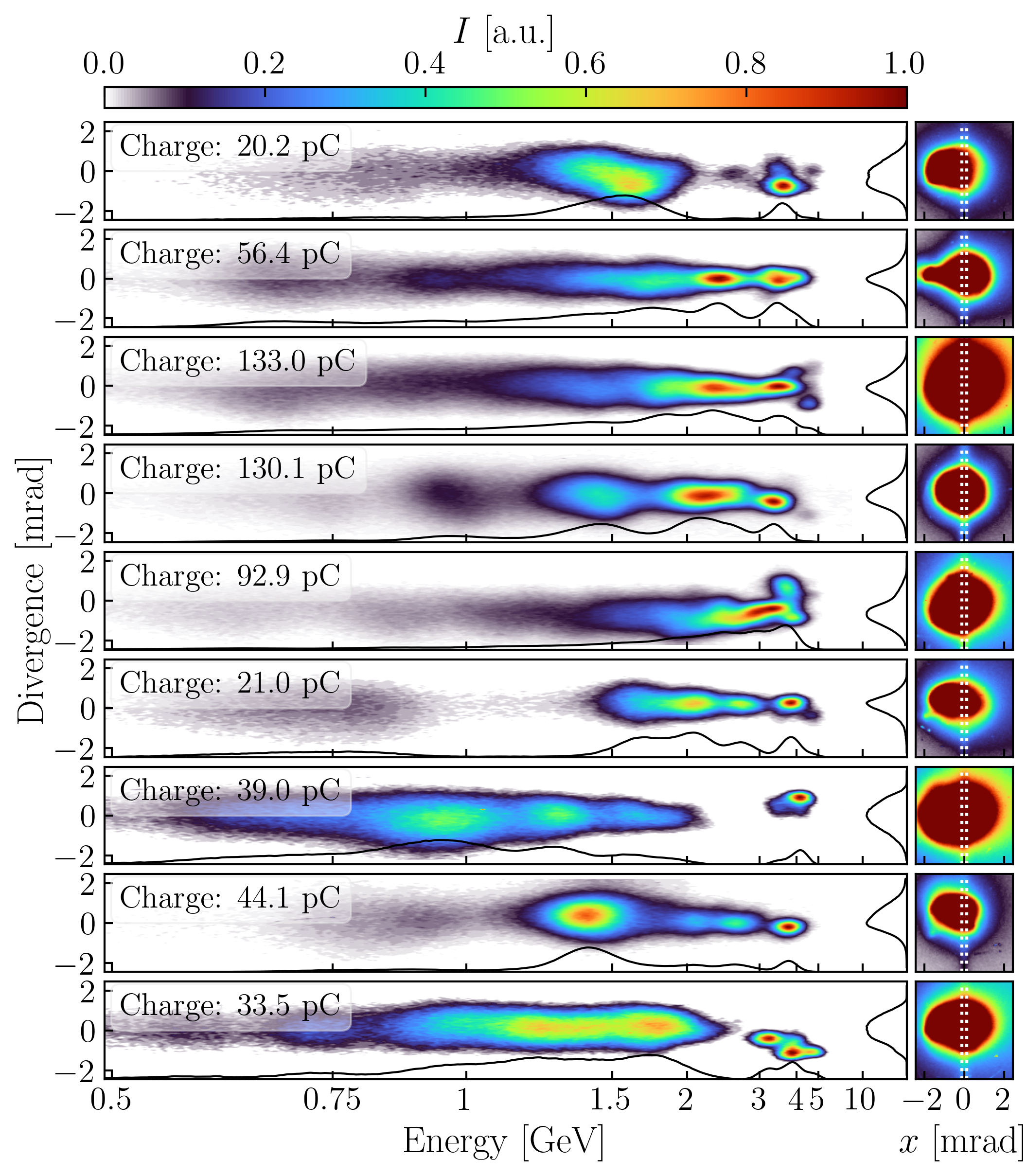}
\caption{Selection of normalized high-energy electron spectra obtained during acceleration in a 20~cm nozzle at 0.2~Hz with a backing pressure of 37–42~bar. Each spectrum is accompanied by a corresponding pointing image, with the 1~mm spectrometer slit indicated by white dashed lines. Black curves at the bottom and right side of each spectrum represent the signal integrated over divergence and energy axes, respectively.
}
\label{fig:heelectrons}
\end{figure}
This behavior originates from the relatively large reservoir volume, designed to suppress turbulence and enable steady-state flow, but leading to a filling time comparable to the valve opening duration. The nozzle was therefore operated in a transient flow regime. Future optimization will focus on reducing the reservoir volume and refining the internal geometry to enable faster filling and a more uniform density profile within the available opening time.

The highest-energy shots were obtained with 7.2~J contained within the $\mathrm{1/e^2}$ focal-spot radius (see Figure~\ref{fig:FSmeas}(a)), for delays from 1.7 to 2.1~ns between the channel-forming pulse and the LWFA drive pulse, at backing pressures of $37–42$~bar. The corresponding average neutral density was measured 8~mm above the nozzle orifice using ``fluorescence method'' \cite{Shrock2022}, which compares the fluorescence of the Bessel focus over the gas jet with known neutral density fluorescence signals generated in helium backfill (measured using a helium-calibrated pressure gauge). The resulting density range is $1.30-1.45\times 10^{18}~$cm$^{-3}$, with standard deviation of $4\times 10^{17}~$cm$^{-3}$ along the gas jet. The fluorescence measurement shown in Figure~\ref{fig:guiding}(i) was performed using the \textit{side-view} diagnostic described in Section~\ref{sec:experimental_setup}, where helium fluorescence was induced by $J_0$ Bessel-beam ionization of the gas sheet.

Under these conditions, a $1\mathrm{\%}$ nitrogen--helium mixture filled the entire gas jet, enabling ionization injection over an extended longitudinal region of the plasma channel. The plasma channel size was controlled by adjusting the delay for hydrodynamic expansion, with optimal guiding occurring when the channel supported a matched mode waist $w_{\mathrm{ch}} \approx w_0 \approx 32~\mu$m. The resulting electron spectra were typically broadband, with a mean electron energy of $\approx 3.5~$GeV and high-energy features extending up to $\sim5~$GeV (see Figure~\ref{fig:heelectrons}). Based on FLUKA \cite{Battistoni2015, Ahdida2022, Donadon2024, Hugo2024} Monte Carlo simulations (version 4-3.4) of the 1~mm slit spectrometer setup, the uncertainty in the reconstructed electron energy is estimated to be $\sim15\mathrm{\%}$ at 3.5 GeV and $\sim20\mathrm{\%}$ at 5 GeV. The maximum observed electron energy is governed by the fraction of laser energy coupled into the plasma waveguide, the resulting effective acceleration length, and the achievable accelerating field under the present laser and plasma conditions. For the measured L3 laser parameters on target ($a_0 = 2.55$) and assuming a plasma electron density of $2.7\times10^{17}~$cm$^{-3}$ (corresponding to a $\sim5\times$ lower on-axis density compared to the measured neutral density on target), the expected accelerating gradient is on the order of 38~GV/m, consistent with wakefield amplitudes reported in similar experiments operating in this parameter regime~\cite{Miao2022}. Ionization injection is most efficient during \textit{Stage II mode beating}~\cite{Shrock2024}, which, under the present channel conditions, occurs $\sim 5$~cm into the gas target, limiting the effective acceleration length to $\sim15$~cm. Using Lu scaling~\cite{Lu2007} with $a_0 = 2.55$, $\tau_{\mathrm{FWHM}} = 30.4~$fs, $w_0 = 32.1~\mu$m, $n_e = 2.7\times10^{17}~$cm$^{-3}$, and a corresponding dephasing length $l_{\mathrm{deph}} = 15$~cm, the expected maximum electron energy is 5.5~GeV, consistent with the measured results.

\begin{figure}[t]
\centering
\includegraphics[width=\columnwidth]{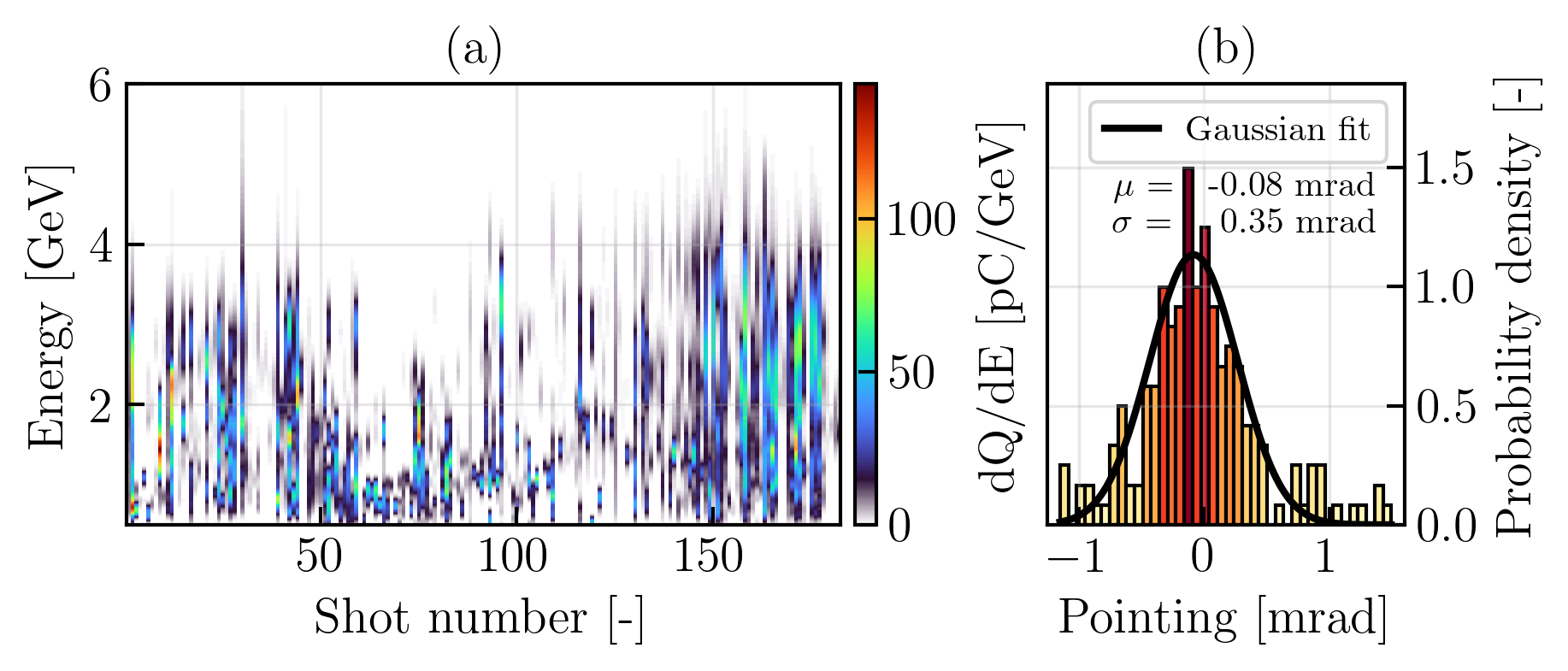}
\caption{\textit{Statistical analysis of electron beam stability.}
(a) Evolution and distribution of the electron beam energy over multiple data sets acquired during optimization.
(b) Statistical distribution of the electron beam pointing along the axis perpendicular to the driver laser polarization, shown as a probability-density histogram with a Gaussian fit.}
\label{fig:electrons-stats}
\end{figure}

Each spectrum in Figure~\ref{fig:heelectrons} is shown together with the corresponding electron-beam pointing image recorded at the entrance of the magnetic spectrometer, where the 1~mm entrance slit is indicated by a dashed line. The accelerated electron beams exhibited characteristic features of self-waveguided propagation, including stable pointing and low divergence, with typical divergences of $1.2 \pm 0.3$~mrad (FWHM) for electron energies above 0.5~GeV and $<0.6$~mrad for electrons exceeding 3~GeV. Shot-to-shot variations in electron beam pointing along the axis perpendicular to the driver laser polarization, inferred from spectra above 0.5~GeV, were $\approx 0.35~$mrad. This value is obtained by analyzing all shots from Figure~\ref{fig:electrons-stats}(a), integrating the charge along the energy axis, and calculating the centroid position for each shot. The resulting distribution is presented as a histogram in Figure~\ref{fig:electrons-stats}(b) and fitted with a normal (Gaussian) distribution, yielding $\mu = -0.08$~mrad and $\sigma = 0.35$~mrad. The electron spectra exhibited multi-peaked structures, consistent with mode-beating-induced ionization injection occurring at multiple longitudinal positions along the plasma channel~\cite{Miao2022, Shrock2024}.

The reported accelerated charge is obtained by integrating the signal measured by the electron spectrometer and therefore represents a lower bound on the total charge, as a significant fraction of the beam does not pass through the slit collimator. The uncertainty in the absolute charge measurement is estimated to be on the order of $15-25\mathrm{\%}$, primarily due to systematic effects in the spectrometer response, energy-dependent detector sensitivity, and secondary particle production. Shot-to-shot variations in charge and spectral shape arise from laser pointing fluctuations characterized by a standard deviation of $\approx 17~\mu$m, which affects laser--channel coupling and leads to variations of $\sim 30\mathrm{\%}$ in guided energy, as well as from injection dynamics and variations in the transverse electron-beam profile. Under otherwise identical experimental conditions, high-energy electron acceleration was achieved in $\sim 40\mathrm{\%}$ of laser shots.

The electron beam energy evolution over multiple data sets acquired during optimization (e.g., delay between pulses, backing pressure, and group delay dispersion (GDD)) is shown in Figure~\ref{fig:electrons-stats}(a). The analysis includes all successfully accelerated electron beams with charge exceeding 0.5~pC transmitted through the spectrometer slit to the detector.

Additional acceleration experiments were carried out using $10\mathrm{\%}$ nitrogen--helium and $5\mathrm{\%}$ argon--helium mixtures to investigate the influence of dopant concentration and ionization potential on the injection dynamics.

\section{Discussion}
The results demonstrate that the all-reflective optical setup with an off-axis axicon provides stable and efficient guiding of high-power femtosecond laser pulses in helium, achieving both high-repetition-rate operation (3.3~Hz) and multi-GeV electron acceleration (0.2~Hz). The use of reflective components throughout the beamline eliminates thermally induced distortions typical of transmissive optics and could enhance stability during extended operation. In particular, the off-axis axicon configuration minimizes back-reflections and enables straightforward alignment---a crucial advantage for high-average-power systems and long-duration experimental campaigns.

Compared with previous implementations of self-wave- guiding and optical-field-ionized plasma channels,  our approach streamlines the self-waveguided LWFA scheme, eliminating the past requirements of additional compressors or multi-beam synchronization. The combination of post-compressor beam splitting and all-reflective optics allows the channel-forming beam to be derived directly from the main pulse without compromising pulse duration or spatial phase quality. This technique thus provides an accessible route toward stable and repeatable plasma guiding, fully compatible with existing user-facility laser architectures.

The observed guiding of petawatt-class laser pulses at 3.3~Hz over centimeter-scale distances confirms the robustness of OFI-generated plasma channels even at high repetition rates, consistent with prior low-energy studies. Furthermore, the electron acceleration experiments at 0.2~Hz demonstrate that this method can produce substantial electron energy gain, confirming its suitability for long plasma channels and higher drive-pulse energies. The use of helium, rather than hydrogen, simplifies gas handling and reduces operational hazards while maintaining efficient ionization and wake excitation.

\section{Conclusion}
We have demonstrated a high-repetition-rate, all-reflective laser guiding and electron acceleration setup employing an off-axis reflective axicon in helium plasma. The system supports stable optical guiding at 3.3 Hz and enables acceleration of electron beams to energies of $\sim5$~GeV at 0.2~Hz, limited by the available laser parameters. This represents the first implementation of self-waveguiding in helium using a fully reflective optical system. The configuration preserves femtosecond pulse duration, minimizes nonlinear distortions, and enhances reproducibility during extended operation.

The simplicity and robustness of this approach make it highly attractive for deployment in user facilities where modifications to the main laser system are constrained. Beyond serving as a compact and efficient method for high-energy electron acceleration, the presented scheme provides a foundation for future high-repetition-rate plasma accelerators and for the development of laser-driven secondary radiation sources, including compact X-ray and positron sources. Future work will focus on extending the plasma channel length, achieving higher-repetition-rate electron acceleration with petawatt-class lasers, and integrating advanced diagnostics and feedback control for real-time optimization of guiding and acceleration.

\begin{center}\textbf{Acknowledgement}\end{center}

\noindent  We sincerely thank the L3 laser team for providing the laser beam and their support during the experiments. We thank John J. Felice for helpful contributions to this study. This work was supported by the National Science Foundation and Czech Science Foundation under NSF-GACR collaborative Grant No.~2206059 from the Czech Science Foundation Grant No. 22-42963L. This work was supported by the U.S. Department of Energy (DE-SC0015516, LaserNetUS DE-SC0019076/ FWP\#SCW1668, and DE-SC0011375), and the Defense Advanced Research Projects Agency (DARPA) under the Muons for Science and Security Program. E.~Rockafellow was supported by NSF GRFP (Grant No. DGE 1840340).

\begin{center}\textbf{References}\end{center}
\begingroup
\bibliographystyle{unsrtnat} 
\renewcommand{\section}[2]{} 
\bibliography{references.tex}

\begin{thebibliography}{73}
\providecommand{\natexlab}[1]{#1}
\providecommand{\url}[1]{\texttt{#1}}
\expandafter\ifx\csname urlstyle\endcsname\relax
  \providecommand{\doi}[1]{doi: #1}\else
  \providecommand{\doi}{doi: \begingroup \urlstyle{rm}\Url}\fi

\bibitem[Tajima and Dawson(1979)]{Tajima1979}
T.~Tajima and J.~M. Dawson.
\newblock Laser electron accelerator.
\newblock \emph{Phys. Rev. Lett.}, 43:\penalty0 267--270, Jul 1979.
\newblock \doi{10.1103/PhysRevLett.43.267}.
\newblock URL \url{https://link.aps.org/doi/10.1103/PhysRevLett.43.267}.

\bibitem[Galletti et~al.(2024)Galletti, Assmann, Couprie, Ferrario, Giannessi,
  Irman, Pompili, and Wang]{Galletti2024}
M.~Galletti, R.~Assmann, M.~E. Couprie, M.~Ferrario, L.~Giannessi, A.~Irman,
  R.~Pompili, and W.~Wang.
\newblock Prospects for free-electron lasers powered by
  plasma-wakefield-accelerated beams.
\newblock \emph{Nature Photonics}, 18\penalty0 (8):\penalty0 780--791, 2024.
\newblock \doi{10.1038/s41566-024-01474-3}.
\newblock URL \url{https://doi.org/10.1038/s41566-024-01474-3}.

\bibitem[Wang et~al.(2021)Wang, Feng, Ke, Yu, Xu, Qi, Chen, Qin, Zhang, Fang,
  Liu, Jiang, Wang, Wang, Yang, Wu, Leng, Liu, Li, and Xu]{Wang2021}
W.~Wang, K.~Feng, L.~Ke, C.~Yu, Y.~Xu, R.~Qi, Y.~Chen, Z.~Qin, Z.~Zhang,
  M.~Fang, J.~Liu, K.~Jiang, H.~Wang, C.~Wang, X.~Yang, F.~Wu, Y.~Leng, J.~Liu,
  R.~Li, and Z.~Xu.
\newblock Free-electron lasing at 27 nanometres based on a laser wakefield
  accelerator.
\newblock \emph{Nature}, 595\penalty0 (7868):\penalty0 516--520, 2021.
\newblock \doi{10.1038/s41586-021-03678-x}.
\newblock URL \url{https://doi.org/10.1038/s41586-021-03678-x}.

\bibitem[André et~al.(2018)André, Andriyash, Loulergue, Labat, Roussel,
  Ghaith, Khojoyan, Thaury, Valléau, Briquez, Marteau, Tavakoli, N’Gotta,
  Dietrich, Lambert, Malka, Benabderrahmane, Vétéran, Chapuis, Ajjouri,
  Sebdaoui, Hubert, Marcouillé, Berteaud, Leclercq, Ajjouri, Rommeluère,
  Bouvet, Duval, Kitegi, Blache, Mahieu, Corde, Gautier, Phuoc, Goddet,
  Lestrade, Herbeaux, Évain, Szwaj, Bielawski, Tafzi, Rousseau, Smartsev,
  Polack, Dennetière, Bourassin-Bouchet, Oliveira, and Couprie]{Andre2018}
T.~André, I.~A. Andriyash, A.~Loulergue, M.~Labat, E.~Roussel, A.~Ghaith,
  M.~Khojoyan, C.~Thaury, M.~Valléau, F.~Briquez, F.~Marteau, K.~Tavakoli,
  P.~N’Gotta, Y.~Dietrich, G.~Lambert, V.~Malka, C.~Benabderrahmane,
  J.~Vétéran, L.~Chapuis, T.~El Ajjouri, M.~Sebdaoui, N.~Hubert,
  O.~Marcouillé, P.~Berteaud, N.~Leclercq, M.~El Ajjouri, P.~Rommeluère,
  F.~Bouvet, J.-P. Duval, C.~Kitegi, F.~Blache, B.~Mahieu, S.~Corde,
  J.~Gautier, K.~Ta Phuoc, J.~P. Goddet, A.~Lestrade, C.~Herbeaux, C.~Évain,
  C.~Szwaj, S.~Bielawski, A.~Tafzi, P.~Rousseau, S.~Smartsev, F.~Polack,
  D.~Dennetière, C.~Bourassin-Bouchet, C.~De Oliveira, and M.-E. Couprie.
\newblock Control of laser plasma accelerated electrons for light sources.
\newblock \emph{Nature Communications}, 9\penalty0 (1):\penalty0 1334, 2018.
\newblock \doi{10.1038/s41467-018-03776-x}.
\newblock URL \url{https://doi.org/10.1038/s41467-018-03776-x}.

\bibitem[Sarri et~al.(2013)Sarri, Schumaker, Di~Piazza, Vargas, Dromey,
  Dieckmann, Chvykov, Maksimchuk, Yanovsky, He, Hou, Nees, Thomas, Keitel,
  Zepf, and Krushelnick]{Sarri2013}
G.~Sarri, W.~Schumaker, A.~Di~Piazza, M.~Vargas, B.~Dromey, M.~E. Dieckmann,
  V.~Chvykov, A.~Maksimchuk, V.~Yanovsky, Z.~H. He, B.~X. Hou, J.~A. Nees,
  A.~G.~R. Thomas, C.~H. Keitel, M.~Zepf, and K.~Krushelnick.
\newblock Table-top laser-based source of femtosecond, collimated,
  ultrarelativistic positron beams.
\newblock \emph{Phys. Rev. Lett.}, 110:\penalty0 255002, Jun 2013.
\newblock \doi{10.1103/PhysRevLett.110.255002}.
\newblock URL \url{https://link.aps.org/doi/10.1103/PhysRevLett.110.255002}.

\bibitem[Sarri et~al.(2015)Sarri, Poder, Cole, Schumaker, Piazza, Reville,
  Dzelzainis, Doria, Gizzi, Grittani, Kar, Keitel, Krushelnick, Kuschel,
  Mangles, Najmudin, Shukla, Silva, Symes, Thomas, Vargas, Vieira, and
  Zepf]{Sarri2015}
G.~Sarri, K.~Poder, J.~M. Cole, W.~Schumaker, A.~Di Piazza, B.~Reville,
  T.~Dzelzainis, D.~Doria, L.~A. Gizzi, G.~Grittani, S.~Kar, C.~H. Keitel,
  K.~Krushelnick, S.~Kuschel, S.~P.~D. Mangles, Z.~Najmudin, N.~Shukla, L.~O.
  Silva, D.~Symes, A.~G.~R. Thomas, M.~Vargas, J.~Vieira, and M.~Zepf.
\newblock Generation of neutral and high-density electron–positron pair
  plasmas in the laboratory.
\newblock \emph{Nature Communications}, 6\penalty0 (1):\penalty0 6747, 2015.
\newblock \doi{10.1038/ncomms7747}.
\newblock URL \url{https://doi.org/10.1038/ncomms7747}.

\bibitem[Corde et~al.(2013)Corde, Ta~Phuoc, Lambert, Fitour, Malka, Rousse,
  Beck, and Lefebvre]{Corde2013}
S.~Corde, K.~Ta~Phuoc, G.~Lambert, R.~Fitour, V.~Malka, A.~Rousse, A.~Beck, and
  E.~Lefebvre.
\newblock Femtosecond x rays from laser-plasma accelerators.
\newblock \emph{Rev. Mod. Phys.}, 85:\penalty0 1--48, Jan 2013.
\newblock \doi{10.1103/RevModPhys.85.1}.
\newblock URL \url{https://link.aps.org/doi/10.1103/RevModPhys.85.1}.

\bibitem[Yan et~al.(2017)Yan, Fruhling, Golovin, Haden, Luo, Zhang, Zhao,
  Zhang, Liu, Chen, Chen, Banerjee, and Umstadter]{Yan2017}
W.~Yan, C.~Fruhling, G.~Golovin, D.~Haden, J.~Luo, P.~Zhang, B.~Zhao, J.~Zhang,
  C.~Liu, M.~Chen, S.~Chen, S.~Banerjee, and D.~Umstadter.
\newblock High-order multiphoton thomson scattering.
\newblock \emph{Nature Photonics}, 11\penalty0 (8):\penalty0 514--520, 2017.
\newblock \doi{10.1038/nphoton.2017.100}.
\newblock URL \url{https://doi.org/10.1038/nphoton.2017.100}.

\bibitem[Mirzaie et~al.(2024)Mirzaie, Hojbota, Kim, Pathak, Pak, Kim, Lee,
  Yoon, Lee, Rhee, Vranic, Amaro, Kim, Sung, and Nam]{Mirzaie2024}
M.~Mirzaie, C.~I. Hojbota, D.~Y. Kim, V.~B. Pathak, T.~G. Pak, C.~M. Kim, H.~W.
  Lee, J.~W. Yoon, S.~K. Lee, Y.~J. Rhee, M.~Vranic, {\'O}.~Amaro, K.~Y. Kim,
  J.~H. Sung, and C.~H. Nam.
\newblock All-optical nonlinear compton scattering performed with a
  multi-petawatt laser.
\newblock \emph{Nature Photonics}, 18\penalty0 (11):\penalty0 1212--1217, 2024.
\newblock \doi{10.1038/s41566-024-01550-8}.
\newblock URL \url{https://doi.org/10.1038/s41566-024-01550-8}.

\bibitem[Cole et~al.(2018)Cole, Behm, Gerstmayr, Blackburn, Wood, Baird, Duff,
  Harvey, Ilderton, Joglekar, Krushelnick, Kuschel, Marklund, McKenna, Murphy,
  Poder, Ridgers, Samarin, Sarri, Symes, Thomas, Warwick, Zepf, Najmudin, and
  Mangles]{Cole2018}
J.~M. Cole, K.~T. Behm, E.~Gerstmayr, T.~G. Blackburn, J.~C. Wood, C.~D. Baird,
  M.~J. Duff, C.~Harvey, A.~Ilderton, A.~S. Joglekar, K.~Krushelnick,
  S.~Kuschel, M.~Marklund, P.~McKenna, C.~D. Murphy, K.~Poder, C.~P. Ridgers,
  G.~M. Samarin, G.~Sarri, D.~R. Symes, A.~G.~R. Thomas, J.~Warwick, M.~Zepf,
  Z.~Najmudin, and S.~P.~D. Mangles.
\newblock Experimental evidence of radiation reaction in the collision of a
  high-intensity laser pulse with a laser-wakefield accelerated electron beam.
\newblock \emph{Phys. Rev. X}, 8:\penalty0 011020, Feb 2018.
\newblock \doi{10.1103/PhysRevX.8.011020}.
\newblock URL \url{https://link.aps.org/doi/10.1103/PhysRevX.8.011020}.

\bibitem[Dreesen et~al.(2014)Dreesen, Green, Browder, Wood, Schwellenbach,
  Ditmire, Tiwari, and Wagner]{Dreesen2014}
W.~Dreesen, J.~A. Green, M.~Browder, J.~Wood, D.~Schwellenbach, T.~Ditmire,
  G.~Tiwari, and C.~Wagner.
\newblock Detection of petawatt laser-induced muon source for rapid high-z
  material detection.
\newblock pages 1--6, 2014.
\newblock \doi{10.1109/NSSMIC.2014.7431088}.

\bibitem[Zhang et~al.(2025)Zhang, Deng, Ge, Wen, Cui, Feng, Wang, Wu, Pan, Liu,
  Deng, Zhang, Chen, Yan, Shan, Yuan, Tian, Qian, Zhu, Xu, Yu, Zhang, Yang,
  Zhou, Gu, Wang, Leng, Sun, and Li]{Zhang2025}
F.~Zhang, L.~Deng, Y.~Ge, J.~Wen, B.~Cui, K.~Feng, H.~Wang, C.~Wu, Z.~Pan,
  H.~Liu, Z.~Deng, Z.~Zhang, L.~Chen, D.~Yan, L.~Shan, Z.~Yuan, C.~Tian,
  J.~Qian, J.~Zhu, Y.~Xu, Y.~Yu, X.~Zhang, L.~Yang, W.~Zhou, Y.~Gu, W.~Wang,
  Y.~Leng, Z.~Sun, and R.~Li.
\newblock Proof-of-principle demonstration of muon production with an
  ultrashort high-intensity laser.
\newblock \emph{Nature Physics}, 21\penalty0 (7):\penalty0 1050--1056, 2025.
\newblock \doi{10.1038/s41567-025-02872-2}.
\newblock URL \url{https://doi.org/10.1038/s41567-025-02872-2}.

\bibitem[Terzani et~al.(2025)Terzani, Kisyov, Greenberg, Le~Pottier, Mironova,
  Picksley, Stackhouse, Tsai, Li, Rockafellow, Miao, Shrock, Heim,
  Garcia-Sciveres, Benedetti, Valentine, Milchberg, Nakamura, Gonsalves, van
  Tilborg, Schroeder, Esarey, and Geddes]{Terzani2025}
D.~Terzani, S.~Kisyov, S.~Greenberg, L.~Le~Pottier, M.~Mironova, A.~Picksley,
  J.~Stackhouse, Hai-En Tsai, R.~Li, E.~Rockafellow, B.~Miao, J.~E. Shrock,
  T.~Heim, M.~Garcia-Sciveres, C.~Benedetti, J.~Valentine, H.~M. Milchberg,
  K.~Nakamura, A.~J. Gonsalves, J.~van Tilborg, C.~B. Schroeder, E.~Esarey, and
  C.~G.~R. Geddes.
\newblock Measurement of directional muon beams generated at the berkeley lab
  laser accelerator.
\newblock \emph{Phys. Rev. Accel. Beams}, 28:\penalty0 103401, Oct 2025.
\newblock \doi{10.1103/kxjr-h7zs}.
\newblock URL \url{https://link.aps.org/doi/10.1103/kxjr-h7zs}.

\bibitem[Poder et~al.(2018)Poder, Tamburini, Sarri, Di~Piazza, Kuschel, Baird,
  Behm, Bohlen, Cole, Corvan, Duff, Gerstmayr, Keitel, Krushelnick, Mangles,
  McKenna, Murphy, Najmudin, Ridgers, Samarin, Symes, Thomas, Warwick, and
  Zepf]{Poder2018}
K.~Poder, M.~Tamburini, G.~Sarri, A.~Di~Piazza, S.~Kuschel, C.~D. Baird,
  K.~Behm, S.~Bohlen, J.~M. Cole, D.~J. Corvan, M.~Duff, E.~Gerstmayr, C.~H.
  Keitel, K.~Krushelnick, S.~P.~D. Mangles, P.~McKenna, C.~D. Murphy,
  Z.~Najmudin, C.~P. Ridgers, G.~M. Samarin, D.~R. Symes, A.~G.~R. Thomas,
  J.~Warwick, and M.~Zepf.
\newblock Experimental signatures of the quantum nature of radiation reaction
  in the field of an ultraintense laser.
\newblock \emph{Phys. Rev. X}, 8:\penalty0 031004, Jul 2018.
\newblock \doi{10.1103/PhysRevX.8.031004}.
\newblock URL \url{https://link.aps.org/doi/10.1103/PhysRevX.8.031004}.

\bibitem[Gonoskov et~al.(2022)Gonoskov, Blackburn, Marklund, and
  Bulanov]{Gonoskov2022}
A.~Gonoskov, T.~G. Blackburn, M.~Marklund, and S.~S. Bulanov.
\newblock Charged particle motion and radiation in strong electromagnetic
  fields.
\newblock \emph{Rev. Mod. Phys.}, 94:\penalty0 045001, Oct 2022.
\newblock \doi{10.1103/RevModPhys.94.045001}.
\newblock URL \url{https://link.aps.org/doi/10.1103/RevModPhys.94.045001}.

\bibitem[Zhang et~al.(2017)Zhang, Hua, Wan, Pai, Guo, Zhang, Ma, Li, Wu, Chu,
  Gu, Xu, Mori, Joshi, Wang, and Lu]{Zhang2017}
C.~J. Zhang, J.~F. Hua, Y.~Wan, C.-H. Pai, B.~Guo, J.~Zhang, Y.~Ma, F.~Li,
  Y.~P. Wu, H.-H. Chu, Y.~Q. Gu, X.~L. Xu, W.~B. Mori, C.~Joshi, J.~Wang, and
  W.~Lu.
\newblock Femtosecond probing of plasma wakefields and observation of the
  plasma wake reversal using a relativistic electron bunch.
\newblock \emph{Phys. Rev. Lett.}, 119:\penalty0 064801, Aug 2017.
\newblock \doi{10.1103/PhysRevLett.119.064801}.
\newblock URL \url{https://link.aps.org/doi/10.1103/PhysRevLett.119.064801}.

\bibitem[Wan et~al.(2023)Wan, Tata, Seemann, Levine, Smartsev, Kroupp, and
  Malka]{Wan2023}
Y.~Wan, S.~Tata, O.~Seemann, E.~Y. Levine, S.~Smartsev, E.~Kroupp, and
  V.~Malka.
\newblock Femtosecond electron microscopy of relativistic electron bunches.
\newblock \emph{Light: Science \& Applications}, 12\penalty0 (1):\penalty0 116,
  2023.
\newblock ISSN 2047-7538.
\newblock \doi{10.1038/s41377-023-01142-1}.
\newblock URL \url{https://doi.org/10.1038/s41377-023-01142-1}.

\bibitem[Russell et~al.(2023)Russell, Campbell, Qian, Cardarelli, Bulanov,
  Bulanov, Grittani, Seipt, Willingale, and Thomas]{Russell2023}
B.~K. Russell, Paul~T. Campbell, Q.~Qian, J.~A. Cardarelli, S.~S. Bulanov,
  S.~V. Bulanov, G.~M. Grittani, D.~Seipt, L.~Willingale, and A.~G.~R. Thomas.
\newblock Ultrafast relativistic electron probing of extreme magnetic fields.
\newblock \emph{Physics of Plasmas}, 30\penalty0 (9):\penalty0 093105, 09 2023.
\newblock ISSN 1070-664X.
\newblock \doi{10.1063/5.0163392}.
\newblock URL \url{https://doi.org/10.1063/5.0163392}.

\bibitem[Labate et~al.(2020)Labate, Palla, Panetta, Avella, Baffigi, Brandi,
  Martino, Fulgentini, Giulietti, K{\"o}ster, Terzani, Tomassini, Traino, and
  Gizzi]{Labate2020}
L.~Labate, D.~Palla, D.~Panetta, F.~Avella, F.~Baffigi, F.~Brandi, F.~Di
  Martino, L.~Fulgentini, A.~Giulietti, P.~K{\"o}ster, D.~Terzani,
  P.~Tomassini, C.~Traino, and L.~A. Gizzi.
\newblock Toward an effective use of laser-driven very high energy electrons
  for radiotherapy: Feasibility assessment of multi-field and intensity
  modulation irradiation schemes.
\newblock \emph{Scientific Reports}, 10\penalty0 (1):\penalty0 17307, 2020.
\newblock \doi{10.1038/s41598-020-74256-w}.
\newblock URL \url{https://doi.org/10.1038/s41598-020-74256-w}.

\bibitem[Guo et~al.(2025)Guo, Liu, Zhou, Liu, Wang, Pi, Wang, Mo, Guo, Hua,
  Wan, and Lu]{Guo2025}
Z.~Guo, S.~Liu, B.~Zhou, J.~Liu, H.~Wang, Y.~Pi, X.~Wang, Y.~Mo, B.~Guo,
  J.~Hua, Y.~Wan, and W.~Lu.
\newblock Preclinical tumor control with a laser-accelerated high-energy
  electron radiotherapy prototype.
\newblock \emph{Nature Communications}, 16\penalty0 (1):\penalty0 1895, 2025.
\newblock ISSN 2041-1723.
\newblock \doi{10.1038/s41467-025-57122-z}.
\newblock URL \url{https://doi.org/10.1038/s41467-025-57122-z}.

\bibitem[Zhou et~al.(2025)Zhou, Guo, Wan, Liu, Peng, Hua, and Lu]{Zhou2025}
B.~Zhou, Z.~Guo, Y.~Wan, S.~Liu, B.~Peng, J.~Hua, and W.~Lu.
\newblock Compact dose delivery of laser-accelerated high-energy electron beams
  toward radiotherapy applications.
\newblock \emph{Phys. Rev. Accel. Beams}, 28:\penalty0 101304, Oct 2025.
\newblock \doi{10.1103/xdsm-7xmf}.
\newblock URL \url{https://link.aps.org/doi/10.1103/xdsm-7xmf}.

\bibitem[Degiovanni et~al.(2016)Degiovanni, Wuensch, and
  Giner~Navarro]{Degiovanni2016}
A.~Degiovanni, W.~Wuensch, and J.~Giner~Navarro.
\newblock Comparison of the conditioning of high gradient accelerating
  structures.
\newblock \emph{Phys. Rev. Accel. Beams}, 19:\penalty0 032001, Mar 2016.
\newblock \doi{10.1103/PhysRevAccelBeams.19.032001}.
\newblock URL
  \url{https://link.aps.org/doi/10.1103/PhysRevAccelBeams.19.032001}.

\bibitem[Leemans et~al.(2006)Leemans, Nagler, Gonsalves, Tóth, Nakamura,
  Geddes, Esarey, Schroeder, and Hooker]{Leemans2006}
W.~P. Leemans, B.~Nagler, A.~J. Gonsalves, Cs. Tóth, K.~Nakamura, C.~G.~R.
  Geddes, E.~Esarey, C.~B. Schroeder, and S.~M. Hooker.
\newblock Gev electron beams from a centimetre-scale accelerator.
\newblock \emph{Nature Physics}, 2\penalty0 (10):\penalty0 696--699, 2006.
\newblock \doi{10.1038/nphys418}.
\newblock URL \url{https://doi.org/10.1038/nphys418}.

\bibitem[Clayton et~al.(2010)Clayton, Ralph, Albert, Fonseca, Glenzer, Joshi,
  Lu, Marsh, Martins, Mori, Pak, Tsung, Pollock, Ross, Silva, and
  Froula]{Clayton2010}
C.~E. Clayton, J.~E. Ralph, F.~Albert, R.~A. Fonseca, S.~H. Glenzer, C.~Joshi,
  W.~Lu, K.~A. Marsh, S.~F. Martins, W.~B. Mori, A.~Pak, F.~S. Tsung, B.~B.
  Pollock, J.~S. Ross, L.~O. Silva, and D.~H. Froula.
\newblock Self-guided laser wakefield acceleration beyond 1 gev using
  ionization-induced injection.
\newblock \emph{Phys. Rev. Lett.}, 105:\penalty0 105003, Sep 2010.
\newblock \doi{10.1103/PhysRevLett.105.105003}.
\newblock URL \url{https://link.aps.org/doi/10.1103/PhysRevLett.105.105003}.

\bibitem[Miao et~al.(2022{\natexlab{a}})Miao, Shrock, Feder, Hollinger,
  Morrison, Nedbailo, Picksley, Song, Wang, Rocca, and Milchberg]{Miao2022}
B.~Miao, J.~E. Shrock, L.~Feder, R.~C. Hollinger, J.~Morrison, R.~Nedbailo,
  A.~Picksley, H.~Song, S.~Wang, J.~J. Rocca, and H.~M. Milchberg.
\newblock Multi-gev electron bunches from an all-optical laser wakefield
  accelerator.
\newblock \emph{Phys. Rev. X}, 12:\penalty0 031038, Sep 2022{\natexlab{a}}.
\newblock \doi{10.1103/PhysRevX.12.031038}.
\newblock URL \url{https://link.aps.org/doi/10.1103/PhysRevX.12.031038}.

\bibitem[Aniculaesei et~al.(2023)Aniculaesei, Ha, Yoffe, Labun, Milton, McCary,
  Spinks, Quevedo, Labun, Sain, Hannasch, Zgadzaj, Pagano, Franco-Altamirano,
  Ringuette, Gaul, Luedtke, Tiwari, Ersfeld, Brunetti, Ruhl, Ditmire, Bruce,
  Donovan, Downer, Jaroszynski, and Hegelich]{Aniculaesei2024}
C.~Aniculaesei, T.~Ha, S.~Yoffe, L.~Labun, S.~Milton, E.~McCary, M.~M. Spinks,
  H.~J. Quevedo, O.~Z. Labun, R.~Sain, A.~Hannasch, R.~Zgadzaj, I.~Pagano,
  J.~A. Franco-Altamirano, M.~L. Ringuette, E.~Gaul, S.~V. Luedtke, G.~Tiwari,
  B.~Ersfeld, E.~Brunetti, H.~Ruhl, T.~Ditmire, S.~Bruce, M.~E. Donovan, M.~C.
  Downer, D.~A. Jaroszynski, and B.~M. Hegelich.
\newblock The acceleration of a high-charge electron bunch to 10 gev in a 10-cm
  nanoparticle-assisted wakefield accelerator.
\newblock \emph{Matter and Radiation at Extremes}, 9\penalty0 (1):\penalty0
  014001, 11 2023.
\newblock ISSN 2468-2047.
\newblock \doi{10.1063/5.0161687}.
\newblock URL \url{https://doi.org/10.1063/5.0161687}.

\bibitem[Schroeder et~al.(2010)Schroeder, Esarey, Geddes, Benedetti, and
  Leemans]{Schroeder2010}
C.~B. Schroeder, E.~Esarey, C.~G.~R. Geddes, C.~Benedetti, and W.~P. Leemans.
\newblock Physics considerations for laser-plasma linear colliders.
\newblock \emph{Phys. Rev. ST Accel. Beams}, 13:\penalty0 101301, Oct 2010.
\newblock \doi{10.1103/PhysRevSTAB.13.101301}.
\newblock URL \url{https://link.aps.org/doi/10.1103/PhysRevSTAB.13.101301}.

\bibitem[Kim et~al.(2013)Kim, Pae, Cha, Kim, Yu, Sung, Lee, Jeong, and
  Lee]{Kim2013}
H.~T. Kim, K.~H. Pae, H.~J. Cha, I~J. Kim, T.~J. Yu, J.~H. Sung, S.~K. Lee,
  T.~M. Jeong, and J.~Lee.
\newblock Enhancement of electron energy to the multi-gev regime by a
  dual-stage laser-wakefield accelerator pumped by petawatt laser pulses.
\newblock \emph{Phys. Rev. Lett.}, 111:\penalty0 165002, Oct 2013.
\newblock \doi{10.1103/PhysRevLett.111.165002}.
\newblock URL \url{https://link.aps.org/doi/10.1103/PhysRevLett.111.165002}.

\bibitem[Morozov et~al.(2018)Morozov, Goltsov, Chen, Scully, and
  Suckewer]{Morozov2018}
A.~Morozov, A.~Goltsov, Q.~Chen, M.~Scully, and S.~Suckewer.
\newblock Ionization assisted self-guiding of femtosecond laser pulses.
\newblock \emph{Physics of Plasmas}, 25\penalty0 (5):\penalty0 053110, 05 2018.
\newblock ISSN 1070-664X.
\newblock \doi{10.1063/1.5021795}.
\newblock URL \url{https://doi.org/10.1063/1.5021795}.

\bibitem[Lemos et~al.(2018)Lemos, Cardoso, Geada, Figueira, Albert, and
  Dias]{Lemos2018}
N.~Lemos, L.~Cardoso, J.~Geada, G.~Figueira, F.~Albert, and J.~M. Dias.
\newblock Guiding of laser pulses in plasma waveguides created by
  linearly-polarized femtosecond laser pulses.
\newblock \emph{Scientific Reports}, 8\penalty0 (1):\penalty0 3165, 2018.
\newblock \doi{10.1038/s41598-018-21392-z}.
\newblock URL \url{https://doi.org/10.1038/s41598-018-21392-z}.

\bibitem[Shalloo et~al.(2018)Shalloo, Arran, Corner, Holloway, Jonnerby,
  Walczak, Milchberg, and Hooker]{Shalloo2018}
R.~J. Shalloo, C.~Arran, L.~Corner, J.~Holloway, J.~Jonnerby, R.~Walczak, H.~M.
  Milchberg, and S.~M. Hooker.
\newblock Hydrodynamic optical-field-ionized plasma channels.
\newblock \emph{Phys. Rev. E}, 97:\penalty0 053203, May 2018.
\newblock \doi{10.1103/PhysRevE.97.053203}.
\newblock URL \url{https://link.aps.org/doi/10.1103/PhysRevE.97.053203}.

\bibitem[Smartsev et~al.(2019)Smartsev, Caizergues, Oubrerie, Gautier, Goddet,
  Tafzi, Phuoc, Malka, and Thaury]{Smartsev2019}
S.~Smartsev, C.~Caizergues, K.~Oubrerie, J.~Gautier, J.-P. Goddet, A.~Tafzi,
  K.~Ta Phuoc, V.~Malka, and C.~Thaury.
\newblock Axiparabola: a long-focal-depth, high-resolution mirror for broadband
  high-intensity lasers.
\newblock \emph{Opt. Lett.}, 44\penalty0 (14):\penalty0 3414--3417, Jul 2019.
\newblock \doi{10.1364/OL.44.003414}.
\newblock URL \url{https://opg.optica.org/ol/abstract.cfm?URI=ol-44-14-3414}.

\bibitem[Shalloo et~al.(2019)Shalloo, Arran, Picksley, von Boetticher, Corner,
  Holloway, Hine, Jonnerby, Milchberg, Thornton, Walczak, and
  Hooker]{Shalloo2019}
R.~J. Shalloo, C.~Arran, A.~Picksley, A.~von Boetticher, L.~Corner,
  J.~Holloway, G.~Hine, J.~Jonnerby, H.~M. Milchberg, C.~Thornton, R.~Walczak,
  and S.~M. Hooker.
\newblock Low-density hydrodynamic optical-field-ionized plasma channels
  generated with an axicon lens.
\newblock \emph{Phys. Rev. Accel. Beams}, 22:\penalty0 041302, Apr 2019.
\newblock \doi{10.1103/PhysRevAccelBeams.22.041302}.
\newblock URL
  \url{https://link.aps.org/doi/10.1103/PhysRevAccelBeams.22.041302}.

\bibitem[Miao et~al.(2020)Miao, Feder, Shrock, Goffin, and Milchberg]{Miao2020}
B.~Miao, L.~Feder, J.~E. Shrock, A.~Goffin, and H.~M. Milchberg.
\newblock Optical guiding in meter-scale plasma waveguides.
\newblock \emph{Phys. Rev. Lett.}, 125:\penalty0 074801, Aug 2020.
\newblock \doi{10.1103/PhysRevLett.125.074801}.
\newblock URL \url{https://link.aps.org/doi/10.1103/PhysRevLett.125.074801}.

\bibitem[Feder et~al.(2020)Feder, Miao, Shrock, Goffin, and
  Milchberg]{Feder2020}
L.~Feder, B.~Miao, J.~E. Shrock, A.~Goffin, and H.~M. Milchberg.
\newblock Self-waveguiding of relativistic laser pulses in neutral gas
  channels.
\newblock \emph{Phys. Rev. Res.}, 2:\penalty0 043173, Nov 2020.
\newblock \doi{10.1103/PhysRevResearch.2.043173}.
\newblock URL \url{https://link.aps.org/doi/10.1103/PhysRevResearch.2.043173}.

\bibitem[Picksley et~al.(2020{\natexlab{a}})Picksley, Alejo, Cowley, Bourgeois,
  Corner, Feder, Holloway, Jones, Jonnerby, Milchberg, Reid, Ross, Walczak, and
  Hooker]{Picksley2020a}
A.~Picksley, A.~Alejo, J.~Cowley, N.~Bourgeois, L.~Corner, L.~Feder,
  J.~Holloway, H.~Jones, J.~Jonnerby, H.~M. Milchberg, L.~R. Reid, A.~J. Ross,
  R.~Walczak, and S.~M. Hooker.
\newblock Guiding of high-intensity laser pulses in 100-mm-long hydrodynamic
  optical-field-ionized plasma channels.
\newblock \emph{Phys. Rev. Accel. Beams}, 23:\penalty0 081303, Aug
  2020{\natexlab{a}}.
\newblock \doi{10.1103/PhysRevAccelBeams.23.081303}.
\newblock URL
  \url{https://link.aps.org/doi/10.1103/PhysRevAccelBeams.23.081303}.

\bibitem[Picksley et~al.(2020{\natexlab{b}})Picksley, Alejo, Shalloo, Arran,
  von Boetticher, Corner, Holloway, Jonnerby, Jakobsson, Thornton, Walczak, and
  Hooker]{Picksley2020b}
A.~Picksley, A.~Alejo, R.~J. Shalloo, C.~Arran, A.~von Boetticher, L.~Corner,
  J.~A. Holloway, J.~Jonnerby, O.~Jakobsson, C.~Thornton, R.~Walczak, and S.~M.
  Hooker.
\newblock Meter-scale conditioned hydrodynamic optical-field-ionized plasma
  channels.
\newblock \emph{Phys. Rev. E}, 102:\penalty0 053201, Nov 2020{\natexlab{b}}.
\newblock \doi{10.1103/PhysRevE.102.053201}.
\newblock URL \url{https://link.aps.org/doi/10.1103/PhysRevE.102.053201}.

\bibitem[Shrock et~al.(2022)Shrock, Miao, Feder, and Milchberg]{Shrock2022}
J.~E. Shrock, B.~Miao, L.~Feder, and H.~M. Milchberg.
\newblock Meter-scale plasma waveguides for multi-gev laser wakefield
  acceleration.
\newblock \emph{Physics of Plasmas}, 29\penalty0 (7):\penalty0 073101, 07 2022.
\newblock ISSN 1070-664X.
\newblock \doi{10.1063/5.0097214}.
\newblock URL \url{https://doi.org/10.1063/5.0097214}.

\bibitem[Shrock et~al.(2024)Shrock, Rockafellow, Miao, Le, Hollinger, Wang,
  Gonsalves, Picksley, Rocca, and Milchberg]{Shrock2024}
J.~E. Shrock, E.~Rockafellow, B.~Miao, M.~Le, R.~C. Hollinger, S.~Wang, A.~J.
  Gonsalves, A.~Picksley, J.~J. Rocca, and H.~M. Milchberg.
\newblock Guided mode evolution and ionization injection in meter-scale
  multi-gev laser wakefield accelerators.
\newblock \emph{Phys. Rev. Lett.}, 133:\penalty0 045002, Jul 2024.
\newblock \doi{10.1103/PhysRevLett.133.045002}.
\newblock URL \url{https://link.aps.org/doi/10.1103/PhysRevLett.133.045002}.

\bibitem[Miao et~al.(2024)Miao, Rockafellow, Shrock, Hancock, Gordon, and
  Milchberg]{Miao2024}
B.~Miao, E.~Rockafellow, J.~E. Shrock, S.~W. Hancock, D.~Gordon, and H.~M.
  Milchberg.
\newblock Benchmarking of hydrodynamic plasma waveguides for multi-gev
  laser-driven electron acceleration.
\newblock \emph{Phys. Rev. Accel. Beams}, 27:\penalty0 081302, Aug 2024.
\newblock \doi{10.1103/PhysRevAccelBeams.27.081302}.
\newblock URL
  \url{https://link.aps.org/doi/10.1103/PhysRevAccelBeams.27.081302}.

\bibitem[Picksley et~al.(2024)Picksley, Stackhouse, Benedetti, Nakamura, Tsai,
  Li, Miao, Shrock, Rockafellow, Milchberg, Schroeder, van Tilborg, Esarey,
  Geddes, and Gonsalves]{Picksley2024}
A.~Picksley, J.~Stackhouse, C.~Benedetti, K.~Nakamura, H.~E. Tsai, R.~Li,
  B.~Miao, J.~E. Shrock, E.~Rockafellow, H.~M. Milchberg, C.~B. Schroeder,
  J.~van Tilborg, E.~Esarey, C.~G.~R. Geddes, and A.~J. Gonsalves.
\newblock Matched guiding and controlled injection in dark-current-free,
  10-gev-class, channel-guided laser-plasma accelerators.
\newblock \emph{Phys. Rev. Lett.}, 133:\penalty0 255001, Dec 2024.
\newblock \doi{10.1103/PhysRevLett.133.255001}.
\newblock URL \url{https://link.aps.org/doi/10.1103/PhysRevLett.133.255001}.

\bibitem[Rockafellow et~al.(2025{\natexlab{a}})Rockafellow, Miao, Shrock,
  Sloss, Le, Hancock, Zahedpour, Hollinger, Wang, King, Zhang, Šišma,
  Grittani, Versaci, Gordon, Williams, Reagan, Rocca, and
  Milchberg]{Rockafellow2025}
E.~Rockafellow, B.~Miao, J.~E. Shrock, A.~Sloss, M.~S. Le, S.~W. Hancock,
  S.~Zahedpour, R.~C. Hollinger, S.~Wang, J.~King, P.~Zhang, J.~Šišma, G.~M.
  Grittani, R.~Versaci, D.~F. Gordon, G.~J. Williams, B.~A. Reagan, J.~J.
  Rocca, and H.~M. Milchberg.
\newblock Development of a high charge 10 gev laser electron accelerator.
\newblock \emph{Physics of Plasmas}, 32\penalty0 (5):\penalty0 053102, 05
  2025{\natexlab{a}}.
\newblock ISSN 1070-664X.
\newblock \doi{10.1063/5.0265640}.
\newblock URL \url{https://doi.org/10.1063/5.0265640}.

\bibitem[Rockafellow et~al.(2025{\natexlab{b}})Rockafellow, Shrock, Miao,
  Sloss, Le, Hancock, Zahedpour, Hollinger, Wang, King, Zhang, Šišma,
  Grittani, Versaci, Gordon, Williams, Reagan, Rocca, and
  Milchberg]{Rockafellow2025b}
E.~Rockafellow, J.~E. Shrock, B.~Miao, A.~J. Sloss, M.~S. Le, S.~W. Hancock,
  S.~Zahedpour, R.~C. Hollinger, S.~Wang, J.~King, P.~Zhang, J.~Šišma, G.~M.
  Grittani, R.~Versaci, D.~F. Gordon, G.~J. Williams, B.~A. Reagan, J.~J.
  Rocca, and H.~M. Milchberg.
\newblock High charge laser acceleration of electrons to 10 gev.
\newblock \emph{Nuclear Instruments and Methods in Physics Research Section A:
  Accelerators, Spectrometers, Detectors and Associated Equipment},
  1077:\penalty0 170586, 2025{\natexlab{b}}.
\newblock ISSN 0168-9002.
\newblock \doi{https://doi.org/10.1016/j.nima.2025.170586}.
\newblock URL
  \url{https://www.sciencedirect.com/science/article/pii/S0168900225003870}.

\bibitem[Bobrova et~al.(2001)Bobrova, Esaulov, Sakai, Sasorov, Spence, Butler,
  Hooker, and Bulanov]{Bobrova2001}
N.~A. Bobrova, A.~A. Esaulov, J.-I. Sakai, P.~V. Sasorov, D.~J. Spence,
  A.~Butler, S.~M. Hooker, and S.~V. Bulanov.
\newblock Simulations of a hydrogen-filled capillary discharge waveguide.
\newblock \emph{Phys. Rev. E}, 65:\penalty0 016407, Dec 2001.
\newblock \doi{10.1103/PhysRevE.65.016407}.
\newblock URL \url{https://link.aps.org/doi/10.1103/PhysRevE.65.016407}.

\bibitem[Butler et~al.(2002)Butler, Spence, and Hooker]{Butler2002}
A.~Butler, D.~J. Spence, and S.~M. Hooker.
\newblock Guiding of high-intensity laser pulses with a hydrogen-filled
  capillary discharge waveguide.
\newblock \emph{Phys. Rev. Lett.}, 89:\penalty0 185003, Oct 2002.
\newblock \doi{10.1103/PhysRevLett.89.185003}.
\newblock URL \url{https://link.aps.org/doi/10.1103/PhysRevLett.89.185003}.

\bibitem[Karsch et~al.(2007)Karsch, Osterhoff, Popp, Rowlands-Rees, Major,
  Fuchs, Marx, Hörlein, Schmid, Veisz, Becker, Schramm, Hidding, Pretzler,
  Habs, Grüner, Krausz, and Hooker]{Karsch2007}
S.~Karsch, J.~Osterhoff, A.~Popp, T.~P. Rowlands-Rees, Z.~Major, M.~Fuchs,
  B.~Marx, R.~Hörlein, K.~Schmid, L.~Veisz, S.~Becker, U.~Schramm, B.~Hidding,
  G.~Pretzler, D.~Habs, F.~Grüner, F.~Krausz, and S.~M. Hooker.
\newblock Gev-scale electron acceleration in a gas-filled capillary discharge
  waveguide.
\newblock \emph{New Journal of Physics}, 9\penalty0 (11):\penalty0 415, nov
  2007.
\newblock \doi{10.1088/1367-2630/9/11/415}.
\newblock URL \url{https://doi.org/10.1088/1367-2630/9/11/415}.

\bibitem[Rowlands-Rees et~al.(2008)Rowlands-Rees, Kamperidis, Kneip, Gonsalves,
  Mangles, Gallacher, Brunetti, Ibbotson, Murphy, Foster, Streeter, Budde,
  Norreys, Jaroszynski, Krushelnick, Najmudin, and Hooker]{RowlandsRees2008}
T.~P. Rowlands-Rees, C.~Kamperidis, S.~Kneip, A.~J. Gonsalves, S.~P.~D.
  Mangles, J.~G. Gallacher, E.~Brunetti, T.~Ibbotson, C.~D. Murphy, P.~S.
  Foster, M.~J.~V. Streeter, F.~Budde, P.~A. Norreys, D.~A. Jaroszynski,
  K.~Krushelnick, Z.~Najmudin, and S.~M. Hooker.
\newblock Laser-driven acceleration of electrons in a partially ionized plasma
  channel.
\newblock \emph{Phys. Rev. Lett.}, 100:\penalty0 105005, Mar 2008.
\newblock \doi{10.1103/PhysRevLett.100.105005}.
\newblock URL \url{https://link.aps.org/doi/10.1103/PhysRevLett.100.105005}.

\bibitem[Gonsalves et~al.(2019)Gonsalves, Nakamura, Daniels, Benedetti,
  Pieronek, de~Raadt, Steinke, Bin, Bulanov, van Tilborg, Geddes, Schroeder,
  T\'oth, Esarey, Swanson, Fan-Chiang, Bagdasarov, Bobrova, Gasilov, Korn,
  Sasorov, and Leemans]{Gonsalves2019}
A.~J. Gonsalves, K.~Nakamura, J.~Daniels, C.~Benedetti, C.~Pieronek, T.~C.~H.
  de~Raadt, S.~Steinke, J.~H. Bin, S.~S. Bulanov, J.~van Tilborg, C.~G.~R.
  Geddes, C.~B. Schroeder, Cs. T\'oth, E.~Esarey, K.~Swanson, L.~Fan-Chiang,
  G.~Bagdasarov, N.~Bobrova, V.~Gasilov, G.~Korn, P.~Sasorov, and W.~P.
  Leemans.
\newblock Petawatt laser guiding and electron beam acceleration to 8 gev in a
  laser-heated capillary discharge waveguide.
\newblock \emph{Phys. Rev. Lett.}, 122:\penalty0 084801, Feb 2019.
\newblock \doi{10.1103/PhysRevLett.122.084801}.
\newblock URL \url{https://link.aps.org/doi/10.1103/PhysRevLett.122.084801}.

\bibitem[Tripathi et~al.(2025)Tripathi, Miao, Sloss, Rockafellow, Shrock,
  Hancock, and Milchberg]{Tripathi2025}
N.~Tripathi, B.~Miao, A.~Sloss, E.~Rockafellow, J.~E. Shrock, S.~W. Hancock,
  and H.~M. Milchberg.
\newblock Longitudinal shaping of plasma waveguides using diffractive axicons
  for laser wakefield acceleration.
\newblock \emph{Opt. Lett.}, 50\penalty0 (10):\penalty0 3441--3444, May 2025.
\newblock \doi{10.1364/OL.561318}.
\newblock URL \url{https://opg.optica.org/ol/abstract.cfm?URI=ol-50-10-3441}.

\bibitem[Durfee and Milchberg(1993)]{Durfee1993}
C.~G. Durfee and H.~M. Milchberg.
\newblock Light pipe for high intensity laser pulses.
\newblock \emph{Phys. Rev. Lett.}, 71:\penalty0 2409--2412, Oct 1993.
\newblock \doi{10.1103/PhysRevLett.71.2409}.
\newblock URL \url{https://link.aps.org/doi/10.1103/PhysRevLett.71.2409}.

\bibitem[Durfee et~al.(1995)Durfee, Lynch, and Milchberg]{Durfee1995}
C.~G. Durfee, J.~Lynch, and H.~M. Milchberg.
\newblock Development of a plasma waveguide for high-intensity laser pulses.
\newblock \emph{Phys. Rev. E}, 51:\penalty0 2368--2389, Mar 1995.
\newblock \doi{10.1103/PhysRevE.51.2368}.
\newblock URL \url{https://link.aps.org/doi/10.1103/PhysRevE.51.2368}.

\bibitem[Corkum et~al.(1989)Corkum, Burnett, and Brunel]{Corkum1989}
P.~B. Corkum, N.~H. Burnett, and F.~Brunel.
\newblock Above-threshold ionization in the long-wavelength limit.
\newblock \emph{Phys. Rev. Lett.}, 62:\penalty0 1259--1262, Mar 1989.
\newblock \doi{10.1103/PhysRevLett.62.1259}.
\newblock URL \url{https://link.aps.org/doi/10.1103/PhysRevLett.62.1259}.

\bibitem[Lorenz et~al.(2019)Lorenz, Grittani, Chacon-Golcher, Lazzarini,
  Limpouch, Nawaz, Nevrkla, Vilanova, and Levato]{Lorenz2019}
S.~Lorenz, G.~Grittani, E.~Chacon-Golcher, C.~M. Lazzarini, J.~Limpouch,
  F.~Nawaz, M.~Nevrkla, L.~Vilanova, and T.~Levato.
\newblock Characterization of supersonic and subsonic gas targets for laser
  wakefield electron acceleration experiments.
\newblock \emph{Matter and Radiation at Extremes}, 4\penalty0 (1):\penalty0
  015401, 01 2019.
\newblock ISSN 2468-2047.
\newblock \doi{10.1063/1.5081509}.
\newblock URL \url{https://doi.org/10.1063/1.5081509}.

\bibitem[Miao et~al.(2025)Miao, Shrock, Rockafellow, Sloss, and
  Milchberg]{Miao2025}
B.~Miao, J.~E. Shrock, E.~Rockafellow, A.~J. Sloss, and H.~M. Milchberg.
\newblock Meter-scale supersonic gas jets for multi-gev laser-plasma
  accelerators.
\newblock \emph{Review of Scientific Instruments}, 96\penalty0 (4):\penalty0
  043003, 04 2025.
\newblock ISSN 0034-6748.
\newblock \doi{10.1063/5.0248959}.
\newblock URL \url{https://doi.org/10.1063/5.0248959}.

\bibitem[Pieronek et~al.(2020)Pieronek, Gonsalves, Benedetti, Bulanov, van
  Tilborg, Bin, Swanson, Daniels, Bagdasarov, Bobrova, Gasilov, Korn, Sasorov,
  Geddes, Schroeder, Leemans, and Esarey]{Pieronek2020}
C.~V. Pieronek, A.~J. Gonsalves, C.~Benedetti, S.~S. Bulanov, J.~van Tilborg,
  J.~H. Bin, K.~K. Swanson, J.~Daniels, G.~A. Bagdasarov, N.~A. Bobrova, V.~A.
  Gasilov, G.~Korn, P.~V. Sasorov, C.~G.~R. Geddes, C.~B. Schroeder, W.~P.
  Leemans, and E.~Esarey.
\newblock Laser-heated capillary discharge waveguides as tunable structures for
  laser-plasma acceleration.
\newblock \emph{Physics of Plasmas}, 27\penalty0 (9):\penalty0 093101, 09 2020.
\newblock ISSN 1070-664X.
\newblock \doi{10.1063/5.0014961}.
\newblock URL \url{https://doi.org/10.1063/5.0014961}.

\bibitem[Miao et~al.(2022{\natexlab{b}})Miao, Feder, Shrock, and
  Milchberg]{Miao2022oe}
B.~Miao, L.~Feder, J.~E. Shrock, and H.~M. Milchberg.
\newblock Phase front retrieval and correction of bessel beams.
\newblock \emph{Opt. Express}, 30\penalty0 (7):\penalty0 11360--11371, Mar
  2022{\natexlab{b}}.
\newblock \doi{10.1364/OE.454796}.
\newblock URL \url{https://opg.optica.org/oe/abstract.cfm?URI=oe-30-7-11360}.

\bibitem[Turner et~al.(2022)Turner, Bulanov, Benedetti, Gonsalves, Leemans,
  Nakamura, van Tilborg, Schroeder, Geddes, and Esarey]{Turner2022}
M.~Turner, S.~S. Bulanov, C.~Benedetti, A.~J. Gonsalves, W.~P. Leemans,
  K.~Nakamura, J.~van Tilborg, C.~B. Schroeder, C.~G.~R. Geddes, and E.~Esarey.
\newblock Strong-field qed experiments using the bella pw laser dual beamlines.
\newblock \emph{The European Physical Journal D}, 76\penalty0 (11):\penalty0
  205, 2022.
\newblock \doi{10.1140/epjd/s10053-022-00535-y}.
\newblock URL \url{https://doi.org/10.1140/epjd/s10053-022-00535-y}.

\bibitem[Boucher et~al.(2018)Boucher, Hoyo, Billet, Pinel, Labroille, and
  Courvoisier]{Boucher2018}
P.~Boucher, J.~Del Hoyo, C.~Billet, O.~Pinel, G.~Labroille, and F.~Courvoisier.
\newblock Generation of high conical angle bessel-gauss beams with reflective
  axicons.
\newblock \emph{Appl. Opt.}, 57\penalty0 (23):\penalty0 6725--6728, Aug 2018.
\newblock \doi{10.1364/AO.57.006725}.
\newblock URL \url{https://opg.optica.org/ao/abstract.cfm?URI=ao-57-23-6725}.

\bibitem[Tong and Lin(2005)]{Tong2005}
X.~M. Tong and C.~D. Lin.
\newblock Empirical formula for static field ionization rates of atoms and
  molecules by lasers in the barrier-suppression regime.
\newblock \emph{Journal of Physics B: Atomic, Molecular and Optical Physics},
  38\penalty0 (15):\penalty0 2593, jul 2005.
\newblock \doi{10.1088/0953-4075/38/15/001}.
\newblock URL \url{https://doi.org/10.1088/0953-4075/38/15/001}.

\bibitem[Alejo et~al.(2022)Alejo, Cowley, Picksley, Walczak, and
  Hooker]{Alejo2022}
A.~Alejo, J.~Cowley, A.~Picksley, R.~Walczak, and S.~M. Hooker.
\newblock Demonstration of kilohertz operation of hydrodynamic
  optical-field-ionized plasma channels.
\newblock \emph{Phys. Rev. Accel. Beams}, 25:\penalty0 011301, Jan 2022.
\newblock \doi{10.1103/PhysRevAccelBeams.25.011301}.
\newblock URL
  \url{https://link.aps.org/doi/10.1103/PhysRevAccelBeams.25.011301}.

\bibitem[Lazzarini et~al.(2019)Lazzarini, Goncalves, Grittani, Lorenz, Nevrkla,
  Valenta, Levato, Bulanov, and Korn]{Lazzarini2019}
C.~M. Lazzarini, L.~V. Goncalves, G.~M. Grittani, S.~Lorenz, M.~Nevrkla,
  P.~Valenta, T.~Levato, S.~V. Bulanov, and G.~Korn.
\newblock Electron acceleration at eli-beamlines: Towards high-energy and
  high-repetition rate accelerators.
\newblock \emph{International Journal of Modern Physics A}, 34\penalty0
  (34):\penalty0 1943010, 2019.
\newblock \doi{10.1142/S0217751X19430103}.
\newblock URL \url{https://doi.org/10.1142/S0217751X19430103}.

\bibitem[Willemsen et~al.(2022)Willemsen, Chaulagain, Havl\'{i}\v{c}kov\'{a},
  Borneis, Ebert, Ehlers, Gauch, Gro{\ss}, Kramer, La\v{s}tovi\v{c}ka, Nejdl,
  Rus, Schrader, Tolenis, Van\v{e}k, Velpula, and Weber]{Willemsen2022}
T.~Willemsen, U.~Chaulagain, I.~Havl\'{i}\v{c}kov\'{a}, S.~Borneis, W.~Ebert,
  H.~Ehlers, M.~Gauch, .~Gro{\ss}, D.~Kramer, T.~La\v{s}tovi\v{c}ka, J.~Nejdl,
  B.~Rus, K.~Schrader, T.~Tolenis, F.~Van\v{e}k, P.~K. Velpula, and S.~Weber.
\newblock Large area ion beam sputtered dielectric ultrafast mirrors for
  petawatt laser beamlines.
\newblock \emph{Opt. Express}, 30\penalty0 (4):\penalty0 6129--6141, Feb 2022.
\newblock \doi{10.1364/OE.452249}.
\newblock URL \url{https://opg.optica.org/oe/abstract.cfm?URI=oe-30-4-6129}.

\bibitem[Zymak et~al.(2024)Zymak, Favetta, Grittani, Lazzarini, Tassielli,
  Grenfell, Goncalves, Lorenz, Slukov{\'a}, Vitha, Versaci, Chacon-Golcher,
  Nevrkla, {\v{S}}i{\v{s}}ma, Antipenkov, {\v{S}}obr, Szuba, Staufer,
  Gr{\"u}ner, Lapadula, Ranieri, Piombino, Hafz, Kamperidis, Papp, Mondal,
  Bakule, and Bulanov]{Zymak2024}
I.~Zymak, M.~Favetta, G.~M. Grittani, C.~M. Lazzarini, G.~Tassielli,
  A.~Grenfell, L.~Goncalves, S.~Lorenz, V.~Slukov{\'a}, F.~Vitha, R.~Versaci,
  E.~Chacon-Golcher, M.~Nevrkla, J.~{\v{S}}i{\v{s}}ma, R.~Antipenkov,
  V.~{\v{S}}obr, W.~Szuba, T.~Staufer, F.~Gr{\"u}ner, L.~Lapadula, E.~Ranieri,
  M.~Piombino, N.~A.~M. Hafz, C.~Kamperidis, D.~Papp, S.~Mondal, P.~Bakule, and
  S.~V. Bulanov.
\newblock Characterization of khz repetition rate laser-driven electron beams
  by an inhomogeneous field dipole magnet spectrometer.
\newblock \emph{Photonics}, 11\penalty0 (12), 2024.
\newblock ISSN 2304-6732.
\newblock \doi{10.3390/photonics11121208}.
\newblock URL \url{https://www.mdpi.com/2304-6732/11/12/1208}.

\bibitem[Auerbach and Karpenko(1994)]{Auerbach1994}
J.~M. Auerbach and V.~P. Karpenko.
\newblock Serrated-aperture apodizers for high-energy laser systems.
\newblock \emph{Appl. Opt.}, 33\penalty0 (15):\penalty0 3179--3183, May 1994.
\newblock \doi{10.1364/AO.33.003179}.
\newblock URL \url{https://opg.optica.org/ao/abstract.cfm?URI=ao-33-15-3179}.

\bibitem[International(2024)]{LightTrans_VirtualLabFusion}
LightTrans International.
\newblock \emph{VirtualLab Fusion User’s Manual}.
\newblock LightTrans, 2024.
\newblock URL
  \url{https://www.lighttrans.com/products-services/learning/users-manual.html}.
\newblock Version accessed: 2024 User Manual.

\bibitem[Iaconis and Walmsley(1998)]{Iaconis1998}
C.~Iaconis and I.~A. Walmsley.
\newblock Spectral phase interferometry for direct electric-field
  reconstruction of ultrashort optical pulses.
\newblock \emph{Opt. Lett.}, 23\penalty0 (10):\penalty0 792--794, May 1998.
\newblock \doi{10.1364/OL.23.000792}.
\newblock URL \url{https://opg.optica.org/ol/abstract.cfm?URI=ol-23-10-792}.

\bibitem[Antonsen and Mora(1995)]{Antonsen1995}
T.~M. Antonsen, Jr. and P.~Mora.
\newblock Leaky channel stabilization of intense laser pulses in tenuous
  plasmas.
\newblock \emph{Phys. Rev. Lett.}, 74:\penalty0 4440--4443, May 1995.
\newblock \doi{10.1103/PhysRevLett.74.4440}.
\newblock URL \url{https://link.aps.org/doi/10.1103/PhysRevLett.74.4440}.

\bibitem[Clark and Milchberg(2000)]{Clark2000}
T.~R. Clark and H.~M. Milchberg.
\newblock Optical mode structure of the plasma waveguide.
\newblock \emph{Phys. Rev. E}, 61:\penalty0 1954--1965, Feb 2000.
\newblock \doi{10.1103/PhysRevE.61.1954}.
\newblock URL \url{https://link.aps.org/doi/10.1103/PhysRevE.61.1954}.

\bibitem[Battistoni et~al.(2015)Battistoni, Boehlen, Cerutti, Chin, Esposito,
  Fassò, Ferrari, Lechner, Empl, Mairani, Mereghetti, Ortega, Ranft, Roesler,
  Sala, Vlachoudis, and Smirnov]{Battistoni2015}
G.~Battistoni, T.~Boehlen, F.~Cerutti, P.~W. Chin, L.~S. Esposito, A.~Fassò,
  A.~Ferrari, A.~Lechner, A.~Empl, A.~Mairani, A.~Mereghetti, P.~G. Ortega,
  J.~Ranft, S.~Roesler, P.~R. Sala, V.~Vlachoudis, and G.~Smirnov.
\newblock Overview of the fluka code.
\newblock \emph{Annals of Nuclear Energy}, 82:\penalty0 10--18, 2015.
\newblock ISSN 0306-4549.
\newblock \doi{https://doi.org/10.1016/j.anucene.2014.11.007}.
\newblock URL
  \url{https://www.sciencedirect.com/science/article/pii/S0306454914005878}.
\newblock Joint International Conference on Supercomputing in Nuclear
  Applications and Monte Carlo 2013, SNA + MC 2013. Pluri- and
  Trans-disciplinarity, Towards New Modeling and Numerical Simulation
  Paradigms.

\bibitem[Ahdida et~al.(2022)Ahdida, Bozzato, Calzolari, Cerutti, Charitonidis,
  Cimmino, Coronetti, D’Alessandro, Donadon~Servelle, Esposito, Froeschl,
  García~Alía, Gerbershagen, Gilardoni, Horváth, Hugo, Infantino, Kouskoura,
  Lechner, Lefebvre, Lerner, Magistris, Manousos, Moryc, Ogallar~Ruiz, Pozzi,
  Prelipcean, Roesler, Rossi, Sabaté~Gilarte, Salvat~Pujol, Schoofs,
  Stránský, Theis, Tsinganis, Versaci, Vlachoudis, Waets, and
  Widorski]{Ahdida2022}
C.~Ahdida, D.~Bozzato, D.~Calzolari, F.~Cerutti, N.~Charitonidis, A.~Cimmino,
  A.~Coronetti, G.~L. D’Alessandro, A.~Donadon~Servelle, L.~S. Esposito,
  R.~Froeschl, R.~García~Alía, A.~Gerbershagen, S.~Gilardoni, D.~Horváth,
  G.~Hugo, A.~Infantino, V.~Kouskoura, A.~Lechner, B.~Lefebvre, G.~Lerner,
  M.~Magistris, A.~Manousos, G.~Moryc, F.~Ogallar~Ruiz, F.~Pozzi,
  D.~Prelipcean, S.~Roesler, R.~Rossi, M.~Sabaté~Gilarte, F.~Salvat~Pujol,
  P.~Schoofs, V.~Stránský, C.~Theis, A.~Tsinganis, R.~Versaci, V.~Vlachoudis,
  A.~Waets, and M.~Widorski.
\newblock New capabilities of the fluka multi-purpose code.
\newblock \emph{Frontiers in Physics}, Volume 9 - 2021, 2022.
\newblock ISSN 2296-424X.
\newblock \doi{10.3389/fphy.2021.788253}.
\newblock URL
  \url{https://www.frontiersin.org/journals/physics/articles/10.3389/fphy.2021.788253}.

\bibitem[Donadon et~al.(2024)Donadon, Hugo, Theis, and Vlachoudis]{Donadon2024}
A.~Donadon, G.~Hugo, C.~Theis, and V.~Vlachoudis.
\newblock Flair3 - recasting simulation experiences with the advanced interface
  for fluka and other monte carlo codes.
\newblock \emph{EPJ Web Conf.}, 302:\penalty0 11005, 2024.
\newblock \doi{10.1051/epjconf/202430211005}.
\newblock URL \url{https://doi.org/10.1051/epjconf/202430211005}.

\bibitem[Hugo et~al.(2024)Hugo, Ahdida, Bozzato, Calzolari, Cerutti,
  Ciccotelli, Cimmino, Devienne, Donadon~Servelle, Dyrcz, Salvatore~Esposito,
  Formento, Froeschl, Garc\'{\i}a~Al\'{\i}a, Gilardoni, Gomes, Horv\'ath,
  Humann, Infantino, Lechner, Lefebvre, Lerner, Lorenzon, Lucsanyi, Magistris,
  Marin, Mazzola, Niang, Nowak, Ogallar~Ruiz, Potoine, Pozzi, Prelipcean,
  Rodin, Roesler, Sabat\'e~Gilarte, Sacristan~Barbero, Salvat~Pujol, Schoofs,
  Serban, Sharankov, Theis, Tisi, Tsinganis, Versaci, Vlachoudis, Waets,
  Widorski, and Zymak]{Hugo2024}
G.~Hugo, C.~Ahdida, D.~Bozzato, D.~Calzolari, F.~Cerutti, A.~Ciccotelli,
  A.~Cimmino, A.~Devienne, A.~Donadon~Servelle, P.~K. Dyrcz,
  L.~Salvatore~Esposito, A.~Formento, R.~Froeschl, R.~Garc\'{\i}a~Al\'{\i}a,
  S.~Gilardoni, A.~Gomes, D.~Horv\'ath, B.~Humann, A.~Infantino, A.~Lechner,
  B.~Lefebvre, G.~Lerner, T.~Lorenzon, D.~Lucsanyi, M.~Magistris, S.~Marin,
  G.~Mazzola, S.~Niang, E.~Nowak, F.~Ogallar~Ruiz, J.-B. Potoine, F.~Pozzi,
  D.~Prelipcean, V.~Rodin, S.~Roesler, M.~Sabat\'e~Gilarte,
  M.~Sacristan~Barbero, F.~Salvat~Pujol, P.~Schoofs, A.-G. Serban,
  I.~Sharankov, C.~Theis, M.~Tisi, A.~Tsinganis, R.~Versaci, V.~Vlachoudis,
  A.~Waets, M.~Widorski, and I.~Zymak.
\newblock Latest fluka developments.
\newblock \emph{EPJ Nuclear Sci. Technol.}, 10:\penalty0 20, 2024.
\newblock \doi{10.1051/epjn/2024023}.
\newblock URL \url{https://doi.org/10.1051/epjn/2024023}.

\bibitem[Lu et~al.(2007)Lu, Tzoufras, Joshi, Tsung, Mori, Vieira, Fonseca, and
  Silva]{Lu2007}
W.~Lu, M.~Tzoufras, C.~Joshi, F.~S. Tsung, W.~B. Mori, J.~Vieira, R.~A.
  Fonseca, and L.~O. Silva.
\newblock Generating multi-gev electron bunches using single stage laser
  wakefield acceleration in a 3d nonlinear regime.
\newblock \emph{Phys. Rev. ST Accel. Beams}, 10:\penalty0 061301, Jun 2007.
\newblock \doi{10.1103/PhysRevSTAB.10.061301}.
\newblock URL \url{https://link.aps.org/doi/10.1103/PhysRevSTAB.10.061301}.

\end{thebibliography}
\endgroup

\end{document}